\documentclass[printer]{aa} % for a referee version
\usepackage{graphicx}
\usepackage{txfonts}
\usepackage[3D]{movie15}
\usepackage{hyperref}
\usepackage[UKenglish]{babel}
\usepackage{url}
\begin{document}
%
 %  \title{3D Maps of the Local ISM from Inversion of Distance-Limited Absorption Measurements\\
 %   I: Testing the Method with Simulated Data}
\title{3D maps of the local ISM from inversion of individual color excess measurements}

   %\subtitle{}

\author{R. Lallement\inst{1}, J.-L. Vergely\inst{2}, B. Valette\inst{3}, L. Puspitarini\inst{1}, L. Eyer\inst{4}, L. Casagrande\inst{5}
 %\thanks{Just to show the usage of the elements in the author field}
 }

\institute{GEPI Observatoire de Paris, CNRS, Universit\'e Paris Diderot, Place Jules Janssen  92190 Meudon,  France\\
\email{rosine.lallement@obspm.fr, lucky.puspitarini@obspm.fr}
\and
ACRI-ST, 260 route du Pin Montard, Sophia Antipolis, France\\
\email{jeanluc.vergely@latmos.ipsl.fr}
\and
Institut des Sciences de la Terre, IRD : UR219, Universit\'e de Savoie, CNRS : UMR5275 LGIT-Savoie, 73376 Le-Bourget-du-Lac,
                France
\and
Observatoire de Gen\`eve, Universit\'e de Gen\`eve, Chemin des Maillettes 51, CH-1290 Sauverny, Switzerland
\and
Research School of Astronomy \& Astrophysics, Mount Stromlo Observatory, The Australian National University, ACT 2611, Australia
}
\date{}

% \abstract{}{}{}{}{} 
% 5 {} token are mandatory
 
  \abstract
{} 
   {%3D tomography of interstellar medium may be useful in may aspects.
  Three-dimensional (3D) maps of the Galactic interstellar matter (ISM) are a potential tool of wide use, however accurate and detailed maps are still lacking. One of the ways to construct the maps is to invert individual distance-limited ISM measurements, a method we have here applied to measurements of stellar color excess  in the optical. 
 }
   {We have assembled color excess data  together with the associated parallax or photometric distances to constitute a catalog of $\simeq$ 23,000 sightlines for stars within 2.5 kpc. The photometric data are taken from Str\"{o}mgren catalogs, the Geneva photometric database, and the Geneva-Copenhagen survey. We also included extinctions derived towards open clusters. We applied, to this color excess dataset, an inversion method based on a regularized Bayesian approach, previously used for mapping at closer distances.}
   {We show the dust spatial distribution resulting from the inversion by means of planar cuts through the differential opacity 3D distribution, and by means of  2D maps of the integrated opacity from the Sun up to various distances. The mapping assigns locations to the nearby dense clouds and represents their distribution at the spatial resolution that is allowed by the dataset properties, i.e. of the order of $\simeq$10 pc close to the Sun and increasing to $\simeq$ 100 pc beyond 1 kpc. Biases towards nearby and/or weakly extincted stars make this dataset particularly appropriate to map the local and neighboring cavities, and to locate faint, extended nearby clouds, both goals that are difficult or impossible with other mapping methods. The new maps reveal a $\simeq$1 kpc wide empty region in the third quadrant in the continuation of the so-called CMa tunnel of the Local Cavity, a cavity that we identify as the Superbubble GSH238+00+09 detected in radio emission maps and that is found to be bounded by the Orion and Vela clouds.  The maps also show an extended narrower tunnel in the opposite direction (\textit{l}$\simeq$70$^{\circ}$) that also extends further the Local Bubble and together with it forms a conspicuous cavity bounded by the main Lup, Sco, Oph, Aql, Lac, Cep, and Tau clouds and OB associations. This chain of cavities and surrounding dense regions constitute the first computed representation of the well known Gould belt/Lindblad ring structures. Finally, almost all off-Plane faint features that appear in 2D dust maps have a counterpart in the 3D maps, providing the dust distribution in nearby tenuous clouds.}
   {}

   \keywords{3D ISM map, inversion method}
\authorrunning{Lallement et al}
\titlerunning{3D extinction maps}
   \maketitle
%
%________________________________________________________________

\section{Introduction}

While emission surveys at various wavelengths are providing increasingly detailed maps of interstellar matter (ISM) in the Galaxy, they lack precise information on the distance to the emitting clouds. In particular, distance assignment based on their radial velocities and a mean Galactic rotation curve leads to a poor 3D description of the ISM in the Sun's vicinity, while most of the medium and high-latitude emissions originate from this vicinity.
%, and the method cannot be applied within 10 $^{\circ}$ or so of the Galactic centre and anti-centre directions where the projected velocity along the line of sight drops to zero. 

Other way to obtain realistic 3D distributions of the ISM is to gather absorption data toward target stars located at known and widely distributed distances, and to invert in some way those line of sight data. Absorption data may be gaseous lines that provide absorbing columns of gaseous species or color excess measurements that provide extinctions and dust columns.   In the case of reddening data, they can be derived in a statistical way from stellar photometric surveys by means of color-color/ color-magnitudes diagrams or stellar population synthesis and Galactic models (in particular the Galactic model of \cite{robin03}),  or they can be based on individual stellar reddening values based on photometric or spectrophotometric measurements and appropriate calibrations  (see \cite{marsh06,marsh09}, \cite{saledrew10}, \cite{gontch12}, \cite{chen13}, \cite{knude10}, \cite{cambre11}, \cite{reis11}, \cite{knude12} for details on various data and methods). In the absence of parallax measurements, photometric distances are derived along with the reddening in a consistent way, {or distances are estimated by comparing the density of foreground stars with models (as in \cite{lombardi10})}. The former methods based on large surveys have the strong advantage of being based on a huge amount of targets, however the achievable spatial resolution is limited due to the required statistics and most of those techniques are not appropriate for mapping at small distance due to the need for zero reddening references. The latter methods based on individual sightlines can potentially produce a higher radial spatial resolution (in principle only limited by the mean radial distance between two target stars), and are appropriate for the solar neighborhood. However, little data are available today, and those maps are still restricted to the Sun vicinity. This situation will hopefully change in future thanks to data from  high-resolution multiplex spectrographs and from the ESA astrometric mission Gaia, and for this reason it is useful to investigate inversion techniques and their application to individual reddening data of increasing number, which is the focus of the present work.

 %Idiot il fait seulement des cartes Star counts provide another way to assign distances to absorbing clouds, and have been successfully used (see \cite{cambre99}) but have the same advantages and limitations.
% and, except for those based on prior Galactic models, can not be applied at small distance due to the need for zero reddening references. 
The first attempt to compute a 3D distribution of the ISM by inverting individual line of sight data was made by \cite{vergely01}, who used compiled data on neutral interstellar (IS) absorbers (NaI and HI) as well as the corresponding Hipparcos parallaxes, and inverted those individual data using a robust tomographic method (\cite{tarantola82}). \cite{2003A&A...411..447L} gathered data for the specific purpose of mapping the so-called Local Bubble or Local cavity, a 100-200 pc-wide cavity around the Sun, and applied the same method to a larger NaI data set (around 1,000 sightlines) with Hipparcos distances. It confirmed the view that there is a cavity deficient in cold and neutral interstellar gas and that the closest dense and cold gas is at $\simeq$ 80 pc  (\cite{cox87}). The inverted map also revealed interstellar tunnels which connect the Local Bubble to surrounding cavities. High latitude sightlines with the smallest absorption are found in chimneys, whose directions are perpendicular to the Gould belt plane. The maps show that the Local Bubble is squeezed by surrounding shells in a complicated pattern. \cite{welsh10} presented 600 pc wide NaI and CaII maps using a  catalog of absorptions towards 1,857 early-type stars located within 800 pc of the Sun. While NaI traces dense and neutral IS, CaII traces both dense neutral and ionized gas. The CaII  exhibit strong spatial similarities to those of their equivalent NaI absorption maps, since the dominant features are the dense cloud associations. More recently, \cite{vergely10} further developed the inversion method, updated the sodium maps and inverted $\simeq$ 6,000 extinction measurements based on stellar Str\"{o}mgren photometry and stars possessing Hipparcos parallaxes. The similarity between the locations of the major dust clouds deduced from the extinction inversion and the gas clouds deduced from NaI within 250 pc has been a first test of the inversion method. Most features, the main opaque regions, the Local Cavity boundaries and the tunnels to surrounding cavities were also derived by \cite{reis11} from Str\"{o}mgren photometry measurements and reddening-distance relationships. See also \cite{reis11} for a general review of solar neighborhood mapping, by means of any method. 

Here we extend the inversion and the mapping to larger distances by making use of more distant stars, and by merging target stars possessing Hipparcos parallaxes and targets with photometric distances. Section 2 describes the data used for the inversion. In the section 3, we briefly describe the improved inversion method and discuss its parameters. 
In the section 4, we present the resulting 3D distributions by means of separate planar cuts. In section 5 we show distance-limited 2D reddening maps based on the inverted 3D distribution and integration up to spheres of various radii. We discuss the results in section 6. 
%
 
%The analysis enables to obtain a reliable 3-D density distribution of the ISM in the solar neighborhood.
%In particular, one of these tunnels, defined by stars at 300 to 600 pc from the Sun showing negligible sodium absorption, connects the well known CMa void (\cite{1985ApJ...296..593G}), which is part of the Local Bubble, with the supershell GSH 238+00+09 (\cite{1998ApJ...498..689H}). 

%__________________________________________________________________
\section{Color excess data and distances}

We started with the  \cite{vergely10} E(b-y) dataset, i.e.  color excesses derived from Str\"{o}mgren photometry for nearby stars possessing Hipparcos parallax measurements. We used here parallax values and associated errors from the most recent analysis of \cite{vanleeuwen07}. The types are distributed between B0 to G2, with 60\% of F-G types. 

We added to this list reddening and photometric distances derived self-consistently from Geneva photometric measurements for a large dataset of B stars (\cite{cram99}, \cite{burcram12}, see the Geneva photometric database). The catalog is composed of early-type stars, O and a large majority of B-type stars. There are 614 targets in common with the Str\"{o}mgren-based dataset, and they have been used to establish the relationship between the Str\"{o}mgren  E(b-y) and the Geneva E(B-V) color excess values. This relationship, already described in \cite{raimond12}  was found to be E(B-V) (Geneva)= (1.585 $\pm$ 0.0129) E(b-y) +(0.0221 $\pm$ 0.0007). Thus, we converted the Geneva E(B-V)s into E(b-y)s following this relationship, then converted all E(b-y)s into Johnson E(B-V)s following the relationship E(B-V)$_{J}$=1.335 E(b-y). In the case of repeating stars, we chose to use the Str\"{o}mgren  and Hipparcos distances. 

 We, then, added the catalog of color excess and distances derived for the target stars from the latest revision of the Geneva-Copenhagen Survey (\cite{nord04, casa11}). The catalog is composed of late-type stars, mainly F and G stars. The reddening estimates are based on the intrinsic color calibration of \cite{olsen88}, which has a stated accuracy of $\simeq$ 0.01 mag. Similarly to the former catalogs, such observational errors might lead to negative values of reddening for nearby stars which have zero or very weak reddening. We removed stars already contained in the two other catalogs. 
  
 Finally, we used color excess measurements towards a series of stellar
clusters, as well as the best estimates of the cluster distances, all from the recent new catalog of \cite{dias12}. 

All extinction measurements were scaled to the Johnson system. 
For the last three sources, we used the photometric distances, even for stars possessing a Hipparcos parallax, since they were derived self-consistently along with the extinction. All those datasets were filtered for known binaries in cases of potentially affected extinction and distance derivations.
We kept for all datasets the negative values of the color excess, in order to avoid any statistical bias at small distance. We then retained from this merged list only those stars fulfilling the following conditions: estimated distance smaller than 2,500 pc, estimated distance to the Plane smaller than 300 pc, E(B-V)$_{J}$ $\geq$ -0.02 (see below), and relative uncertainties on the distances smaller than 35\% ({for both Hipparcos and photometric distances)}. The limitation in distances is guided by experience. Keeping a few isolated very distant stars with uncertain distance provides weak additional constraints resulting in elongated structures of poor interest (see the map descriptions). {On the other hand using a threshold on the relative uncertainty allows to maintain moderately distant targets that have a particularly good accuracy, and to exclude those nearby stars that have very large uncertainties.} Note that the limitation in height above the Plane results in a limited distance for the inversion that is a function of the latitude, more precisely d$_{limit}$=300/sin(\textit{b}).  We will come back to this point in section 4. Our final list contains 22,883 remaining targets: 5,106,  12,120, 4,830 and 827 targets are from the four sources respectively. 
%__________________________________________________________________

\section{The inversion method applied to extinction data}

{The present inversion is based on the pioneering work of \cite{tarantola82} who derived general formalisms for non linear least squares inverse problems. We use here their specific formalism adapted to the solution for functions of continuous variables. In our context the differential reddening (or equivalently differential opacity) is a continuous function of the 3D interstellar space.  Given our datasets of distance-limited data, reconstructing such a 3D map is by far under-constrained. However the inversion can be regularized and performed by imposing that the solution is smooth and by using a Bayesian formulation, i.e. the solution is based not only on the color excess data but also on a prior knowledge of the opacity distribution (see \cite{vergely10}). These two information sources complement each other: where the constraints from the data are insufficient, the inversion restores the prior density; in the contrary case, it favors the information contained in the data. For more details on the inversion technique applied to IS line-of-sight data see \cite{vergely10}. We will mention hereafter the specific changes from this previous work.}

\subsection{{Data and associated errors}}

{The data are the color excess measurements towards the target stars and the target distances. It is important to determine their biases and standard deviations. The color excess is assumed to be proportional to the dust opacity towards the target column. At variance with the previous inversion, we make use of independent color excess datasets that are not based on the same wavelength intervals, while still using as a variable a unique differential color excess.  More specifically, during the computation it is defined as the Johnson E(B-V)$_{J}$ color excess per parsec (in units of mag.pc$^{-1}$). Because a color excess is a combination of both the dust volume density and of the dust properties, i.e. of the local reddening law, having merged the data after a simple scaling and using a unique variable is equivalent to implicitly assuming that spatial variabilities in the reddening law do not change significantly the ratios between the various color excess quantities. However, the wavelength ranges being very close, this is a reasonable assumption.}

\subsubsection{Color excess data}
 For all data we carefully kept all meaningful negative values in order to avoid biases at low reddening. We also proceeded in the following conservative way for errors. The standard deviation of the color excess data comes essentially from the uncertainty on the photometric measurements and uncertainties linked to the calibration or intrinsic dispersion in the color-magnitude diagrams. 
Uncertainties on the color excess measurements introduced in the inversion code are as follows: 
i) For the Str\"{o}mgren dataset, uncertainties on the E(b-y)s  were taken from Table 2 of \cite{vergely10}, and correspond to combined errors on the diagram dispersion and on photometry. The errors on E(b-y) were converted to errors on E(B-V) according to the relationship quoted above.  ii) The Geneva database provides a precise determination of the photometric error, which is small and well below the error linked to the use of calibration laws. We used here a conservative value of 0.014 mag for the uncertainty on E(B-V) . iii) In the case of the Geneva-Copenhagen data we estimated conservatively the uncertainties, taken here to be very similarly of the order of 0.016. iv) For the uncertainties on the cluster data, they were taken from the \cite{dias12} catalog that are also conservatively estimated.
{An additional error $\sigma_{cal}=0.01$ is introduced (see also below)  to represent uncertainties linked to the use of different photometric systems and calibration methods. The choice of this value is somewhat arbitrary, and is a compromise between the offset of 0.02 found at the origin of the linear relationship between the Str\"{o}mgren and Geneva values (the most different systems here), and zero. This offset is corrected for, but we do not have any comparison between the cluster data and the other systems, and there are potential departures from linearity that remain hidden within the intrinsic uncertainties for all systems. Also, we deliberately wanted the total error estimate to be conservative.}

\subsubsection{Distances}
% If the extinction data come from different calibration processes, it is needed to apply a cross correlation of the measurement values. 

Uncertainties on the distances are crucial for the inversion process. Relative errors on the distances are either the Hipparcos errors in case of parallax measurements, or arbitrarily taken to be of 20\% for all photometric determinations, except for  the cluster distances for which \cite{dias12} provided conservative estimates. Importantly, in case of Hipparcos or cluster data, as mentioned above, we removed all targets for which the relative error exceeds 35\%. This is important for Hipparcos parallaxes, since it is well known that the accuracy may strongly drop for distant stars ($\gtrsim$ 300 pc). Also, errors on photometric distances to the clusters are not independent, which may introduce stronger discrepancies compared to groups of independent stars in the same region. 

\subsubsection {Propagated errors}

During the inversion, quantities that are computed and adjusted are integrated quantities along path-lengths and as a consequence errors on those integrals arise from both extinction and distance uncertainties. For this reason they are not treated independently during the inversion (see \cite{vergely10}).  In the present work the total errors $\sigma_{tot}$ on the integrated opacities towards the targets have been taken to be the quadratic sum of three terms: 

$\sigma_{tot}=\sqrt{\sigma_{phot}^2+\sigma_{cal}^2+\sigma_{Ed}^2}$ where

 i) $\sigma_{phot}$ is the extinction measurement error quoted above,
 
 ii)  $\sigma_{cal}$ is an additional error equal to 0.01 to represent uncertainties linked to the use of different photometric systems and calibration methods {(see above)}.
 
  iii) $\sigma_{Ed}$ is the error on the distance d propagated to the extinction E(B-V), estimated to be 
$\sigma_{Ed}=E(B-V)\frac{\sigma_d}{d}$ under the assumption of a constant opacity along the LOS. We also assume that  $\frac{\sigma_d}{d}=\frac{\sigma_\pi}{\pi}$  for Hipparcos parallaxes $\pi$, 

{About the assumption that $\sigma_{Ed}=E(B-V)\frac{\sigma_d}{d}$, it may appear highly questionable to assume a constant opacity given the strong clumpiness of the ISM. However, the combined errors are used in the model computation, in the frame of  which the IS matter is much more smoothly distributed than in the actual medium. This makes the assumption significantly more realistic. Moreover, as d increases, $\frac{\sigma_d}{d}$ increases but the assumption generally does not become more invalid, since the smoothness and structure size also increase with distance as a result of the target scarcity.  Finally, if there are strong incompatibilities between some data due to this assumption, uncertainties are accordingly increased during the iteration process (see the Appendix).}

%%%%%%%%%Figure Levy Flight's Illustration of 50000 steps%%%%%%%%%%%%
%\begin{figure}
%\centering
%\includegraphics[width=0.5\linewidth]{Illustration_levyflightstechnique.eps}
%\caption{An illustration of l\'evy flights technique in xy-plane which allowing small and long jumps.}
%\label{levyillust}
%\end{figure}
%%%%%%%%%%%%%%%%%%%%%%%%%%%%%%%%%%%%%%%%%%%%

%%%%%%%%%simulated 3D distribution data%%%%%%%%%%%%%%%%%%%%
%__________________________________________________________________
 \begin{figure*}
   \centering
   \includegraphics[width=13cm]{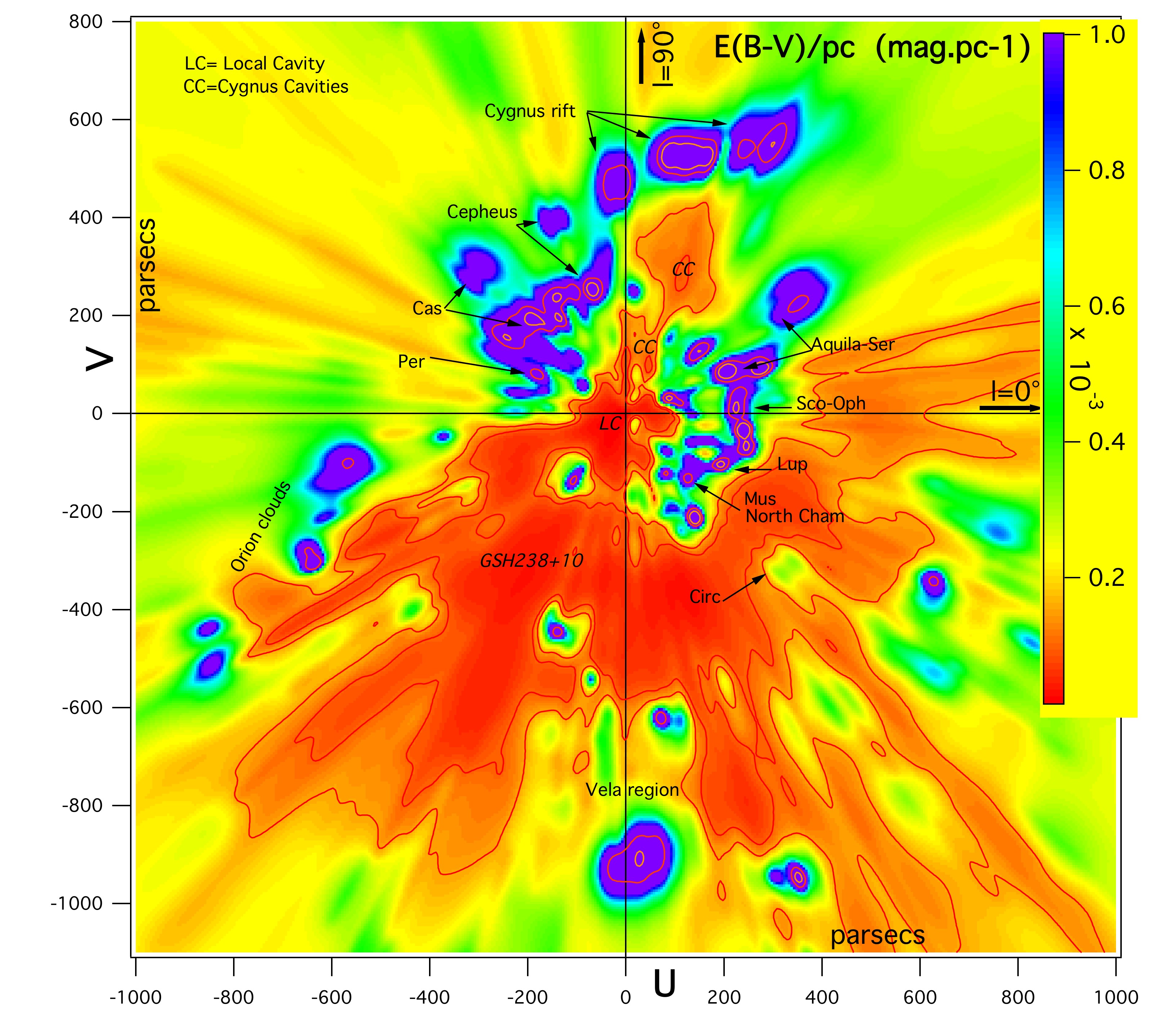}
   \caption{The inverted differential opacity distribution in the Galactic Plane. The Sun is at coordinates (x,y)=(0,0) and the Galactic center direction at right. The opacity increases from red to violet (scale at right). Note the 1,000 pc long cavity centered at \textit{l}$\simeq+225^{\circ}$ that we identified as the major super-bubble GSH238+00+09 detected in radio by \cite{heiles98}. We have called \textit{Cygnus cavities} the empty regions at \textit{l}$\simeq+70^{\circ}$. Several well known opaque regions are indicated on the maps, their identification is based on the literature. The gradient axis for interstellar Helium ionization point to the center of the GSH238+00+09.}
              \label{Figgalplane}%
    \end{figure*}

   \begin{figure}
   \centering
   \includegraphics[width=\linewidth]{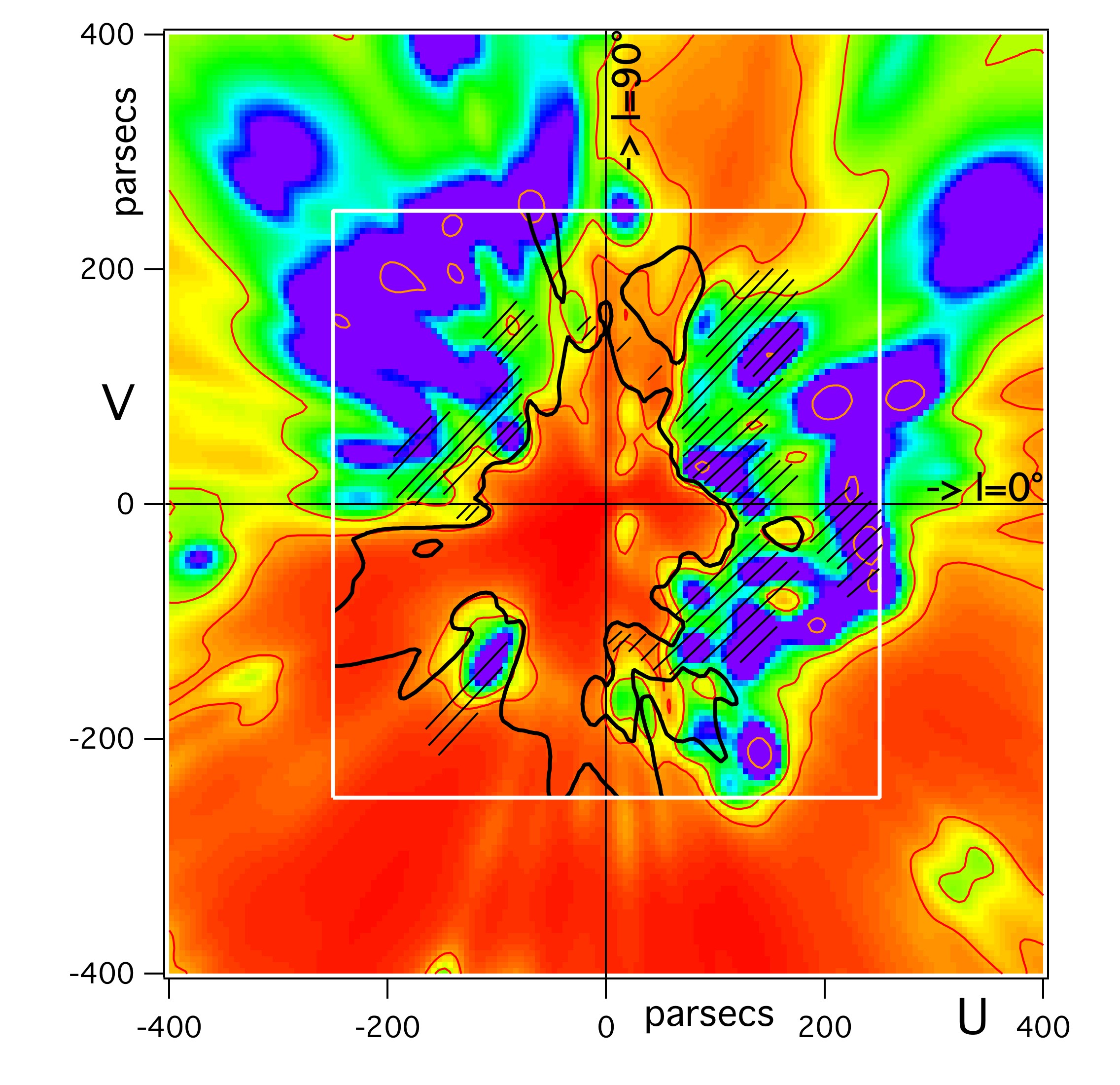}
   \caption{Comparison between the extended map and the 500 pc wide map from \cite{vergely10}. The black line is an iso-contour at dE(B-V)/dr = 0.0002 mag per pc drawn from the previous 3D distribution and delimiting the local cavity. Hatched areas correspond to the closest dense regions revealed in those previous maps. Depending on locations, the new maps reveal the full extent of those opaque regions at larger radial distances, or their division into distinct groups of clouds that the poorer resolution did not allow to distinguish earlier. }
              \label{Figoldnew}%
    \end{figure}

   \begin{figure}
   \centering
   \includegraphics[width=\linewidth]{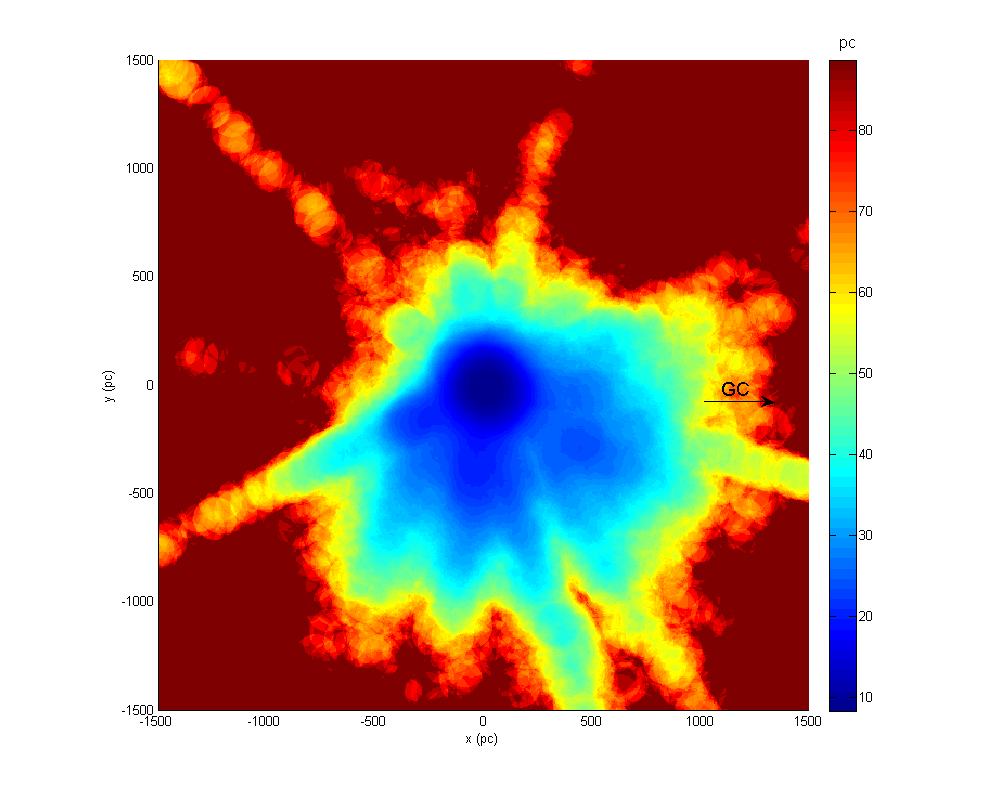}
   \caption{{Achievable resolution in the Galactic plane resulting from the target distribution.}}
              \label{Figresolution}%
    \end{figure}

 \subsection{Prior information on the model and associated variances}

Here the model parameters are functions. We define the correlation kernel $\psi(x,x')$ as the correlation of the opacity between two points in space $x$ and $x'$. It is of fundamental importance since it  controls the spatial variability of the model, the shape of allowed structures and the smoothing length. Evidently the smoothing length cannot be shorter than the average distance between the targets, here varying between a few pc to $\simeq$ 100 pc depending on the distance from the Sun and the location, otherwise the problem is too under-constrained. 

% The correlation kernel enters the covariance operator 
%$C_mf(x)=\int_V\sigma(x)\sigma(x')\psi(x,x')f(x')dV(x')$,

%During the inversion process, the correlation kernel $\psi(x,x')$ controls the smoothing through the size and shape of the favored structures. $\sigma^2(x)$ is the model variance at point $x$ that controls the departures from the prior distribution and 
Making use of a simple Gaussian kernel, 

$\psi(x,x')=\exp(\frac{-\|x-x'\|^2}{2\xi_0^2})$, 

where $\xi_0$ is the smoothing length, yields a poor fit that we interpret as the consequence of  the ISM clearly showing different characteristic lengths. It is thus well-advised to use multi-scale kernels (\cite{serbanjacobsen2001}, \cite{vergely10}). \cite{vergely10} used two exponential kernels, the first one characterizing the warm diffuse matter and the second one the more compact clouds. However, note that the smallest structures (pc and sub-pc scale) cannot be represented because the spatial fluctuations that are smaller than the smoothing length are not detected.

{After a number of tests, we have chosen the following two kernels as best representing the different scales that can be detected given the dataset. They have two different functional forms to produce both an exponential wing and a Gaussian core.   The exponential law is appropriate in case of multi-scale structures and allows to reproduce the intercloud phase and large scales. On the other hand the gaussian kernel allows to reach denser structures, while introducing a threshold in their sizes to avoid too much power in non realistic very small and dense structures. 
The final choice finally has been based on the quality of the adjustments. The kernel is expressed as:}

\begin{equation}
  \begin{array}{l}
  \sigma(x)\sigma(x')\psi(x,x')=\sigma_0(x)\sigma_0(x')\frac{1}{\cosh(-x/\xi_0)}\\
+ \sigma_1(x)\sigma_1(x')\exp(\frac{-x^2}{\xi_1^2}) 
\end{array}
\label{eqkernel}
\end{equation}

{where $\sigma_{(0;1)}^2(x)$ represents the model variance at point $x$ that controls the departures from the prior distribution and $\xi_{(0;1)}$ are the characteristic scales allowed in the model.}

In addition, the density of IS matter decreases strongly with the distance to the Galactic Plane. Hence, the prior model is characterized by an exponential decrease with the distance to the Plane $\rho_{ref}(r,b)=\rho_0\exp(\frac{-|r\sin(b)|}{h_0})$. Note that, following \cite{vergely10}, we have chosen to use $h_{0}$= 200 pc for the opacity scale height, a value deliberately above the average measurement (\cite{chen98}) to avoid the loss of the tenuous high latitude structures during the inversion, which may happen when the prior density is too small. Finally, the computed variable is the logarithm of the opacity per distance $\alpha$ (i.e. $\rho(x)= \exp(\alpha(x))$) to ensure the positivity of the solution.

%Due to the limited number of line-of-sight, the problem is under-constrained, and assumptions must be done under the form of a smoothing length, or equivalently a correlation between the volume opacities at two neighbouring data points.
%Here we have tested the inversion algorithms in the case of two co-existing correlation
%distances. The more general case of multi-scale ($n>2$) kernels is in progress. 

\section{Results}

The line of sight data are inverted to produce a tridimensional differential opacity distribution in a Sun-centered volume whose total dimensions are 4 x 4 x 0.6 kpc$^{3}$, with 0.6 kpc being the vertical extent, i.e. 300 pc above and below the horizontal plane containing the Sun (note that for simplicity we call Galactic Plane this horizontal plane, despite the non null distance of the Sun to the actual Plane). However, only for about half of this volume is the model significantly constrained, the target stars being too sparse at large distance where the opacity distribution remains equal or very close to the prior model, and for this reason the results presented below are restricted to the central regions.  In order to better illustrate the limits of the inversion a \textit{b}=0$^{\circ}$ planar cut in the full inverted volume is shown in the Appendix (Fig A1). We also show illustrate in an Appendix the distribution of stars that are close to the Plane and constrain the distribution. The reason for maintaining targets as distant as
 2,500 pc during the inversion is that  they may contribute to the constraints at closer distance, e.g. in case of very low reddening, and they may reveal some trends. 
The inverted quantity is the smoothed reddening per distance, here E(B-V) in magnitude per pc (color scale in Fig \ref{Figgalplane}). The map can be used to estimate the total reddening towards a target contained in the represented plane, by using the color scale and the Sun-target trace on the map. 
For clarity of the maps we have added a small number of annotations. We recall that (mainly distant) regions where the distribution looks fully homogeneous and varies as a function of Z only are those for which the prior distribution has been unchanged, due to the lack of constraining targets. {We use the Galactic Plane map to discuss the uncertainties associated to the inversion method.}

\subsection{Extinction in the Galactic Plane }

Figure \ref{Figgalplane} shows the color-coded differential opacity distribution in the Plane. We have added iso-opacity contours in order to enhance the characteristics of the distribution. We warn the reader again  that due to the use of the correlation kernel high values of the differential opacity that would correspond to the densest structures and cloud cores are nowhere reached, instead the inverted quantity corresponds to an average over regions of size of the order of 15 or 30 pc. Still, the color-coded map allows to estimate the reddening up to a given distance. 

It is informative to compare this updated map with the former map of \cite{vergely10} based on about a quarter of the present dataset, and we present such a comparison for the Galactic Plane. This is illustrated in Fig \ref{Figoldnew} where we have superimposed on the new map an iso-differential opacity contour computed from the 500 pc wide, former map. The differential opacity value that has been chosen for this iso-contour, namely dE(B-V)/dr = 0.0002 mag per pc corresponds to the marked transition between the Local Cavity and its boundaries. It can be seen that this contour is quite similar to the new Local Cavity boundary now found in the inner part of the new map. The new boundary is traced by the first new contour (thin red line) that has been drawn for dE(B-V)/dr = 0.00016 mag per pc. Hatched areas show the locations of the dense clouds that came out from the previous inversion. It can be seen that, while those areas still correspond to dense regions in the new, more extended maps, the clouds now often extend to larger distances. The reason is that regions beyond the first opaque clouds were simply not showing up previously due to the lack of constraining target stars, those stars being too extinguished, hence absent from the dataset. Instead the prior distribution was kept in those external regions. These limitations still exist for the new maps, albeit now pushed away at larger distances. This is why we call the attention on the fact that dense structures at large distance and located exactly beyond  other, closer dense clouds may be underestimated or even missing in the maps.

{In order to quantify the limitations due to the limited number of distant targets, we show in Fig. \ref{Figresolution} the achievable resolution at each location in the Plane resulting from the target distributions. There are strong asymetries between the quadrants that reflect the mentioned biases and the predominance of empty or dense areas. } We also display in Fig B1 (resp. B2) of the Appendix  those stars that are within 10 (resp 150) pc from the Galactic Plane and are the main contributors to the opacity pattern, superimposed on the Galactic Plane map itself. This comparison allows to figure out at which distance and in which direction constraints are getting too loose and the prior solution is preferred. It is also a way to figure out the minimum size of the structures that can be reconstructed during the inversion, from the distance between the targets. It can be seen that the minimum size increases from $\simeq$ 10 pc close to the Sun (see Fig B1) up to  $\simeq$ 150 pc at 1 kpc (see Fig B2). {Errors on the locations of the clouds depend on the target distribution and evidently distance uncertainties, however they are difficult to quantify. If a structure is defined by a statistically significant number of targets, errors on the distances average out and the center of the structure is correctly defined. The extent of the clouds however increases as a result of all distance uncertainties. In case  only few targets define a structure, there may be very different errors among the various situations. If the targets are angularly close but have different distances, the model produces radially elongated structures that are easily identified, and the radial size of those structures allow to infer the error on the cloud location. If  targets are both scarce and irregularly located towards a structure, e.g. if they are missing towards the densest area, then individual  errors on distances may have their strongest impact, i.e. the error on the cloud location may reach the mean error on the target distances. Fortunately this is not the case for most of the clouds}.

The map reveals the top or bottom parts of the series of dense structures that bound the so-called Local Cavity, the $\simeq$ 100 pc wide empty region around the Sun: the Aquila, Ophiuchus, Scorpius, Lupus, Crux and Centurus dense clouds in the first and fourth quadrant, the Cassiopeia,  Lacerta, Perseus, Taurus, and Orion clouds in the anti-center area. 
It is beyond the scope of this article to discuss in details those clouds, instead we only superimpose on the series of vertical maps presented in the next section the major cloud complexes and OB associations. 

The distribution in Fig \ref{Figgalplane} shows a conspicuous, huge empty cavity in the third quadrant. This cavity is in the continuation of the so-called  CMa {\it tunnel}, the rarefied region that extends up to 130-150 pc in the direction of the star $\epsilon$CMa. A dense region located at $\simeq$ 180 pc marks a partial limit between the two cavities, but it is angularly limited. As shown in other planar cuts, this huge cavity is not limited to the Plane but is extending both above and below to large distances. We note that the existence of this cavity has been inferred by \cite{heiles98} based on radio maps and other emission data. The schematic representation of GSH238+00+09 by \cite{heiles98} corresponds quite well to the geometry that is coming out from the inversion. As already noted in this work, this super-bubble is bounded by the Orion clouds at lower longitude, and by Vela clouds at $\simeq$ 260-270 $^{\circ}$. 

%Consequences on the scenarios of Gould belt formation etc..

%Fig: map of the limiting size of structures (JL: a moins que tu ne veuilles pas le mettre la, mais dans un papier plus theforique)
%description of the main structure

\begin{figure*}%[h!]
\centering
\begin{minipage}[t]{0.7\linewidth}
	\includegraphics[width=\linewidth]{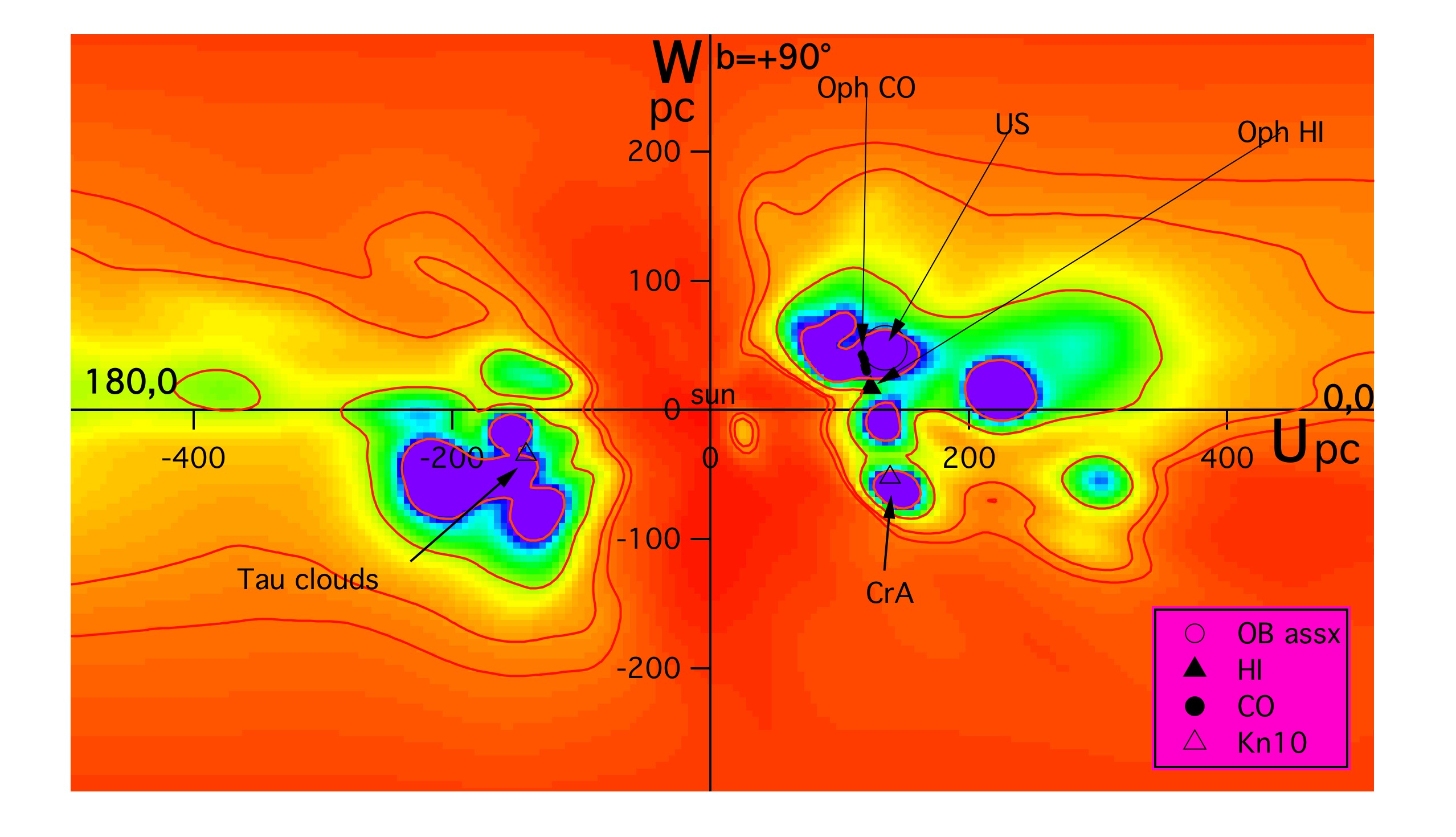}
	\includegraphics[width=\linewidth]{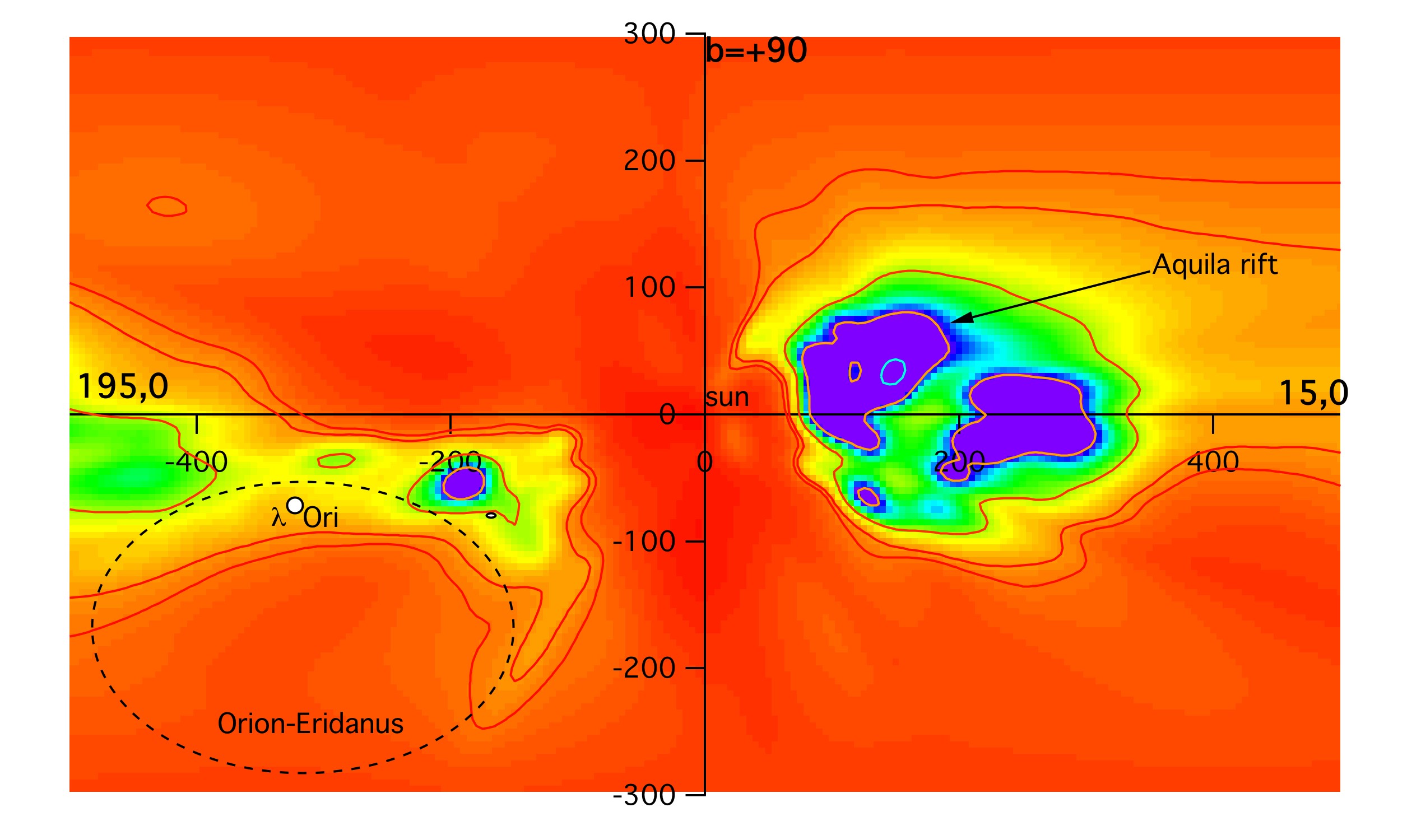}
	\includegraphics[width=\linewidth]{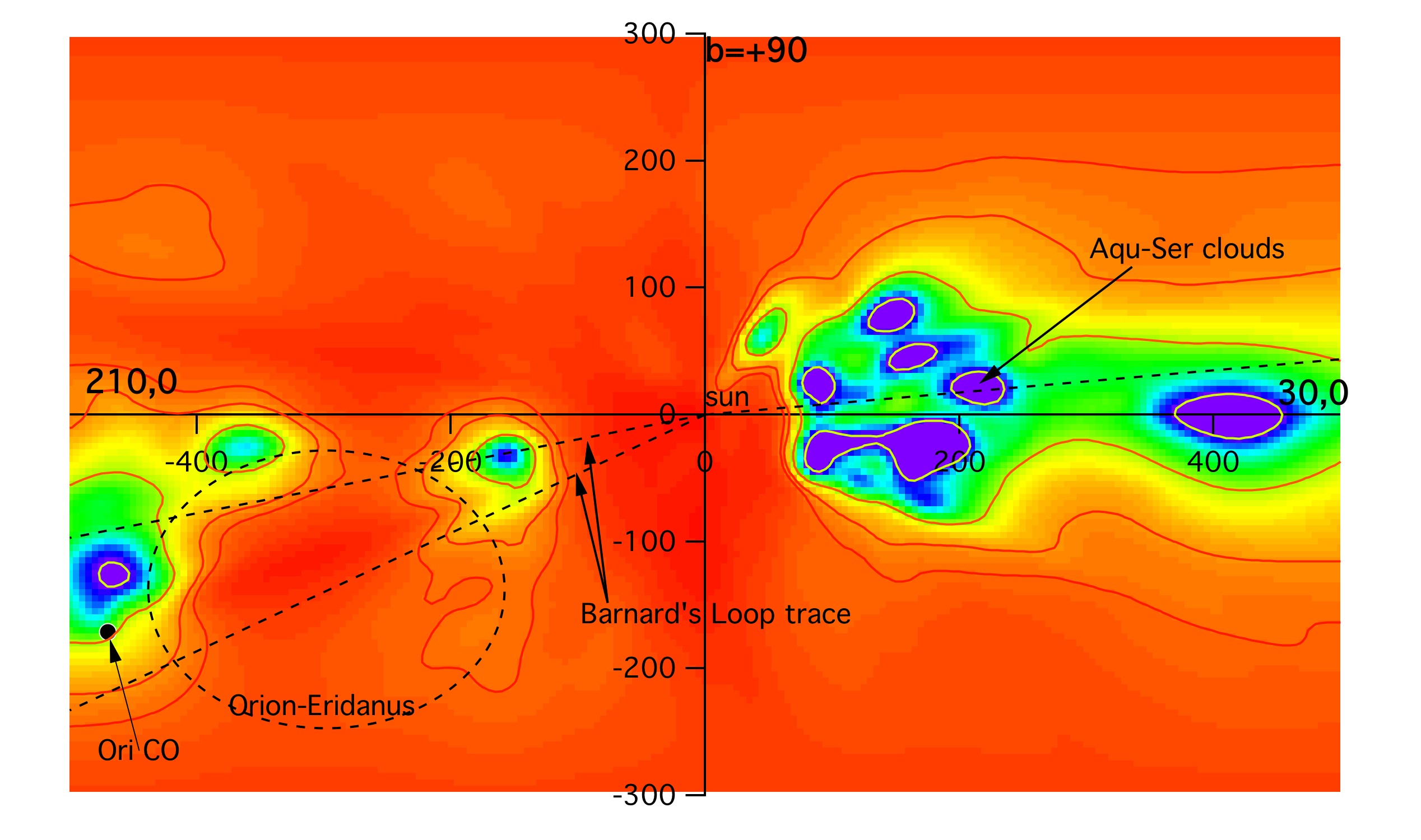}
\end{minipage}\hfill
\caption{Opacity distribution in vertical planes containing the Sun, equally spaced by 15$^{\circ}$. The North pole direction b=+90$^{\circ}$ is at top and longitudes of intersections with the Galactic Plane are indicated. Some iso-differential opacity contours have been superimposed to help visualizing the low and high opacity regions. OB associations from De Zeeuw (1999) as well as CO and HI clouds listed by \cite{perrot03} have been displayed when they are within 25 pc from this vertical plane. We also add the molecular cloud locations derived by \cite{knude10}. Note the tenuous cloud close to the North Pole direction.}
\label{fourvertplanes1}
\end{figure*}

\begin{figure*}
\centering
\begin{minipage}[t]{0.7\linewidth}
	\includegraphics[width=\linewidth]{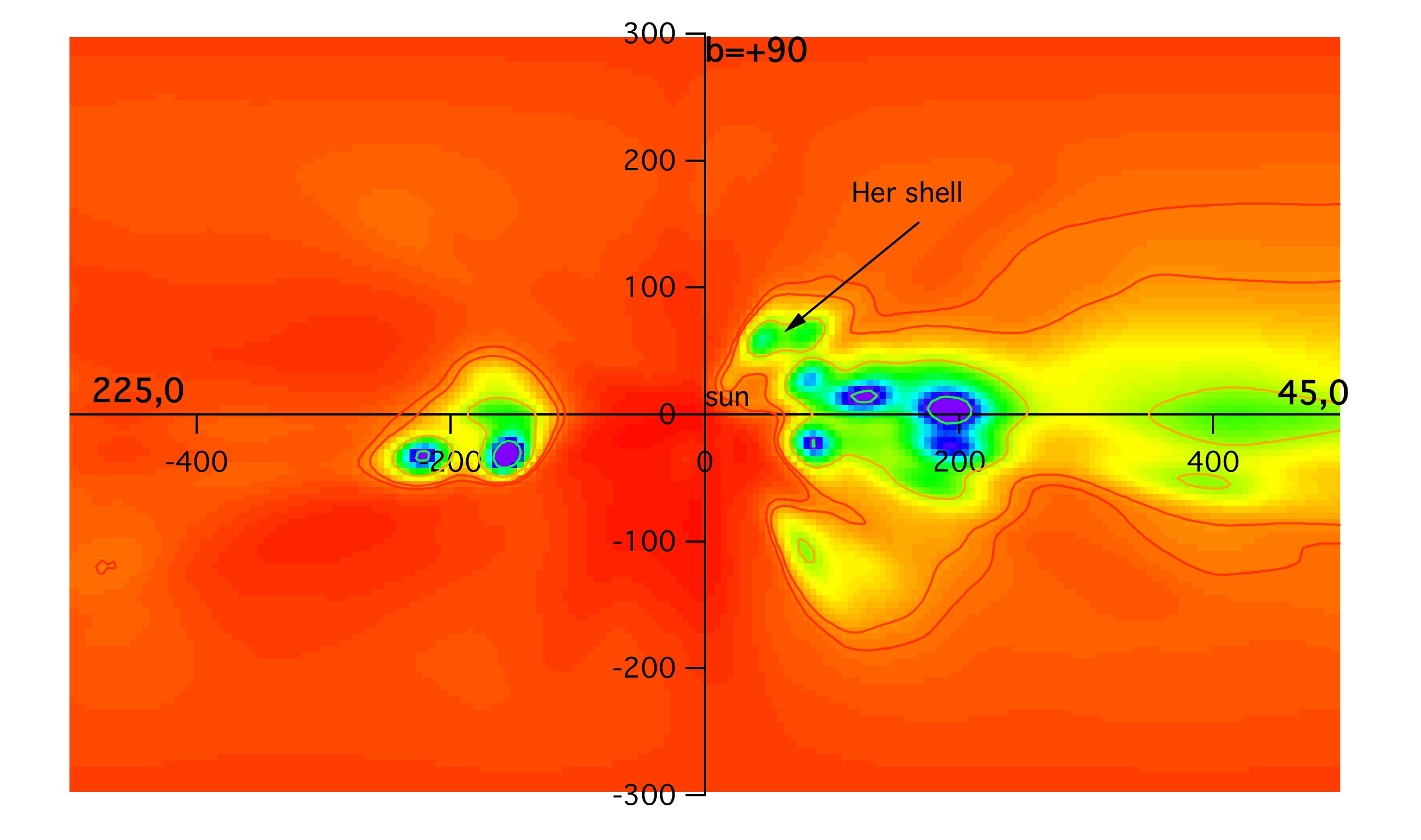}
	\includegraphics[width=\linewidth]{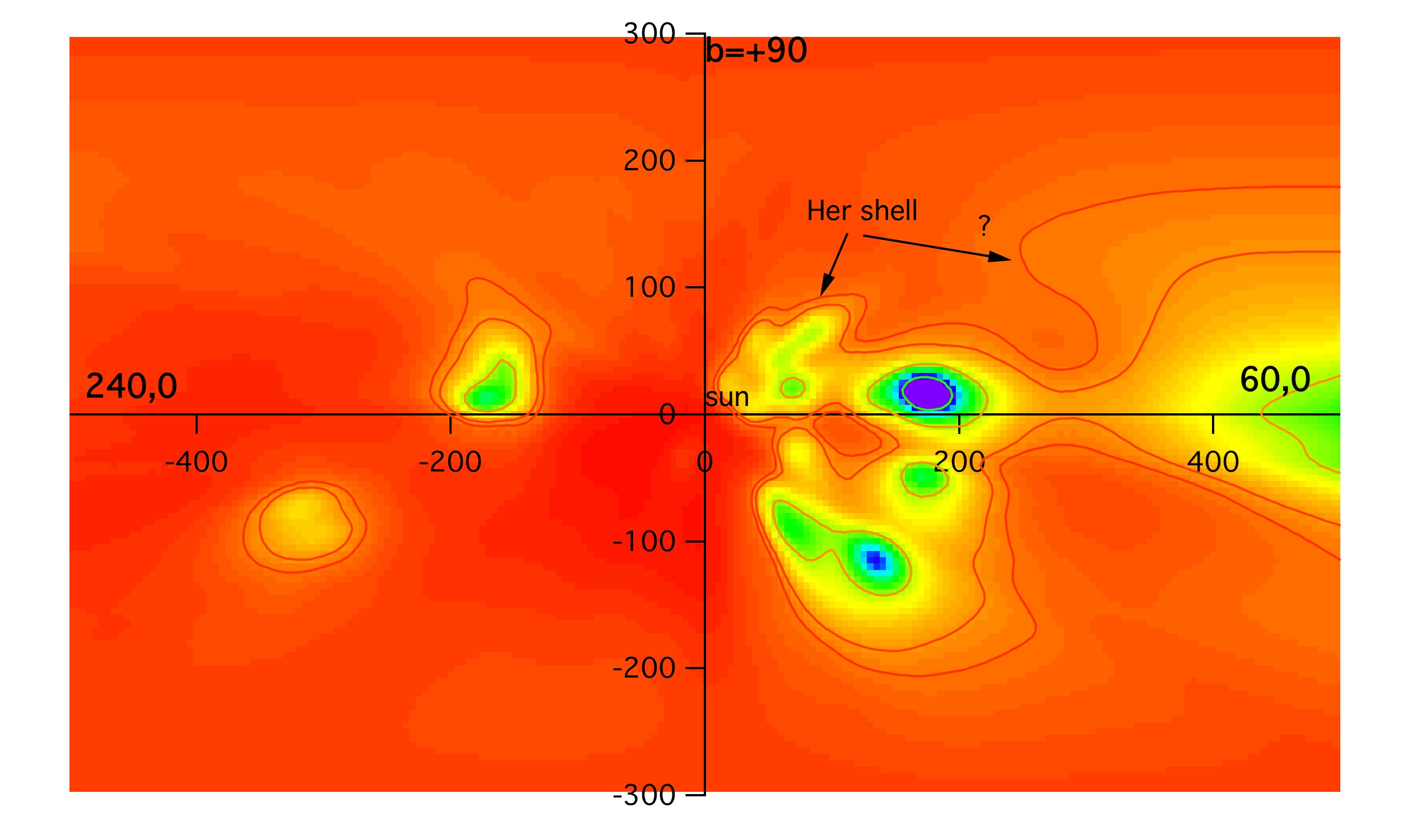}
	\includegraphics[width=\linewidth]{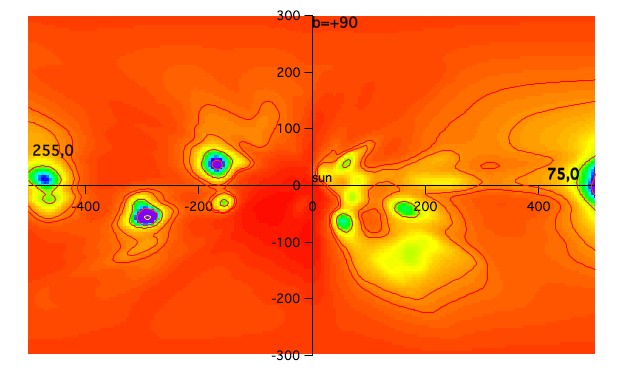}
\end{minipage}\hfill
\caption{Same as fig \ref{fourvertplanes1} for the next three vertical planes. In the \textit{l}=45$^{\circ}$ and \textit{l}=60$^{\circ}$ half-planes (top and middle), the HI Hercules shell is indicated at the  60-150 pc distance derived by \cite{lili92} for its low velocity component. Such a location is in good agreement with the elongated feature at \textit{b}=$\simeq$40$^{\circ}$. The second component at 250 pc may correspond to the fainter distant feature in the \textit{l}=60$^{\circ}$ half-plane.}
\label{fourvertplanes2}
\end{figure*}

\begin{figure*}
\centering
\begin{minipage}[t]{0.7\linewidth}
	\includegraphics[width=\linewidth]{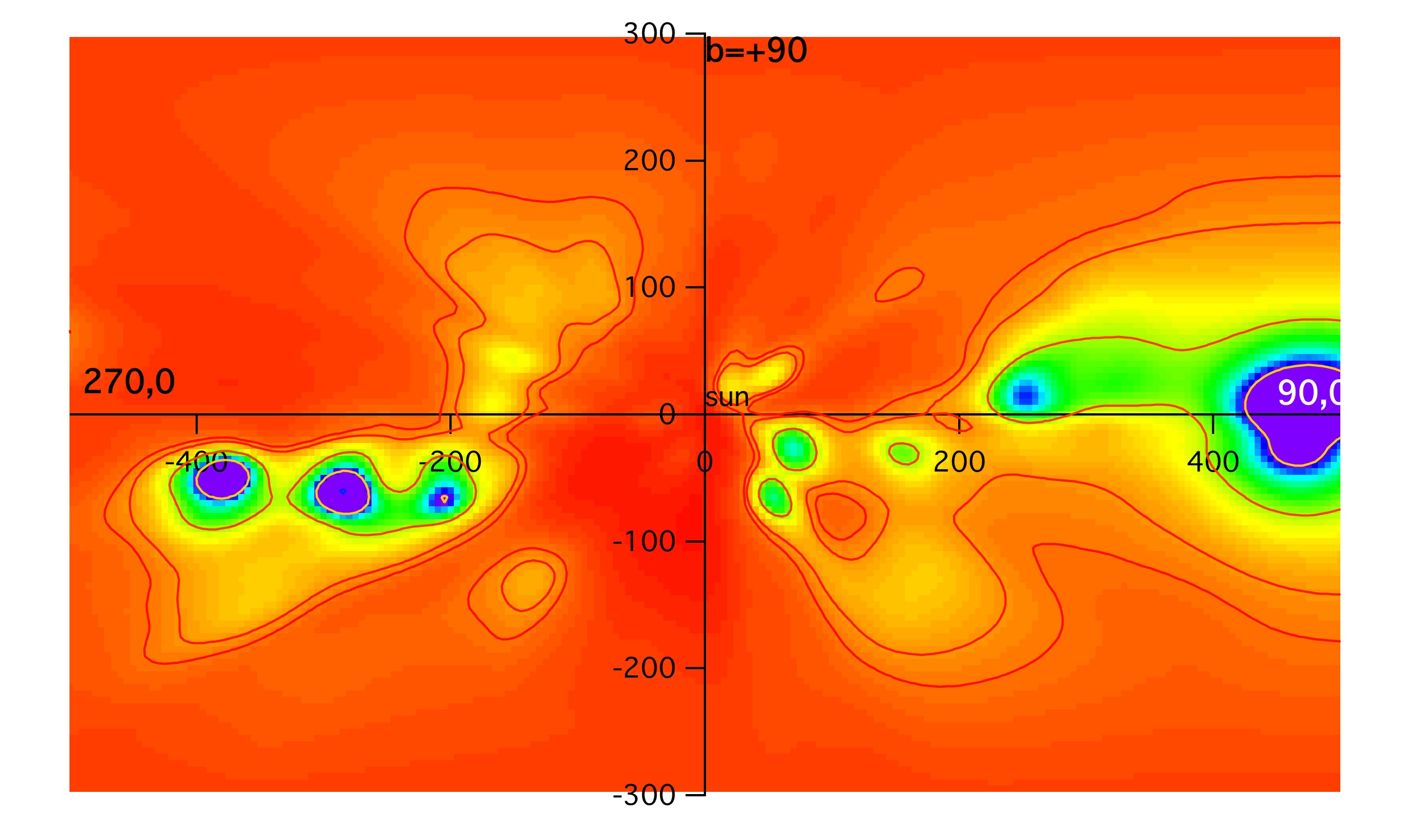}
	\includegraphics[width=\linewidth]{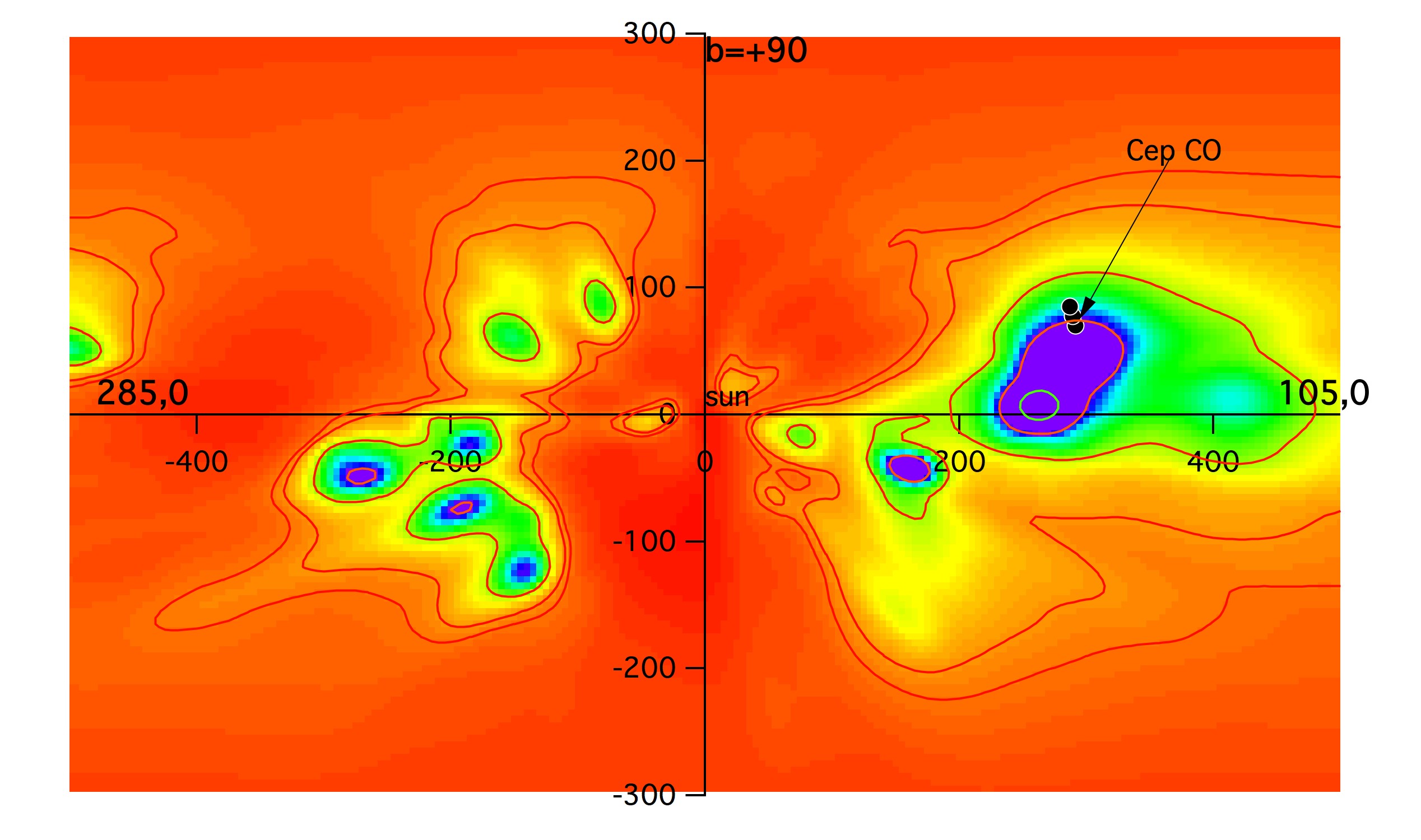}
	\includegraphics[width=\linewidth]{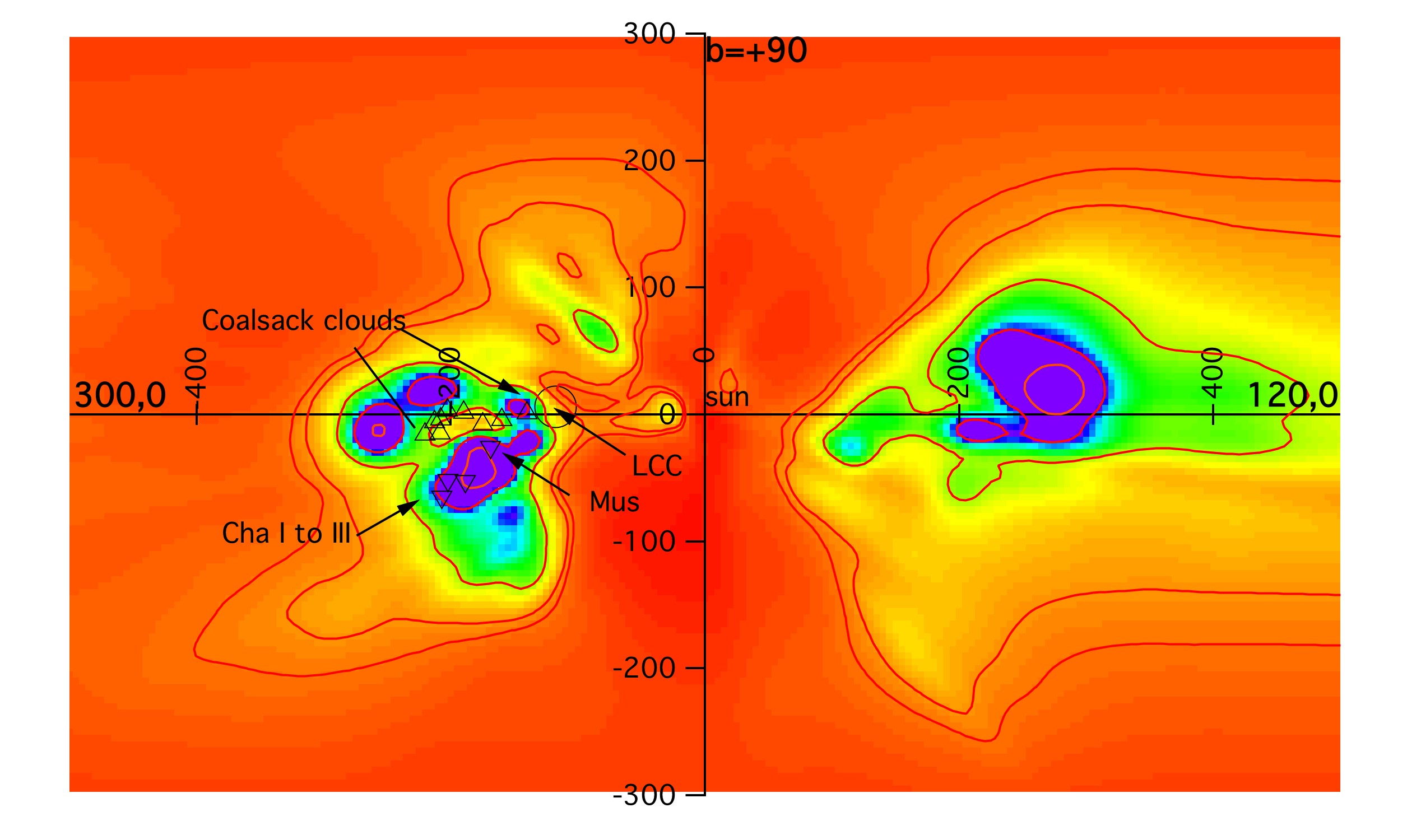}
\end{minipage}\hfill
\caption{Same as fig \ref{fourvertplanes1} for the next three vertical planes.}
\label{fourvertplanes3}
\end{figure*}

\begin{figure*}
\centering
\begin{minipage}[t]{0.7\linewidth}
	\includegraphics[width=\linewidth]{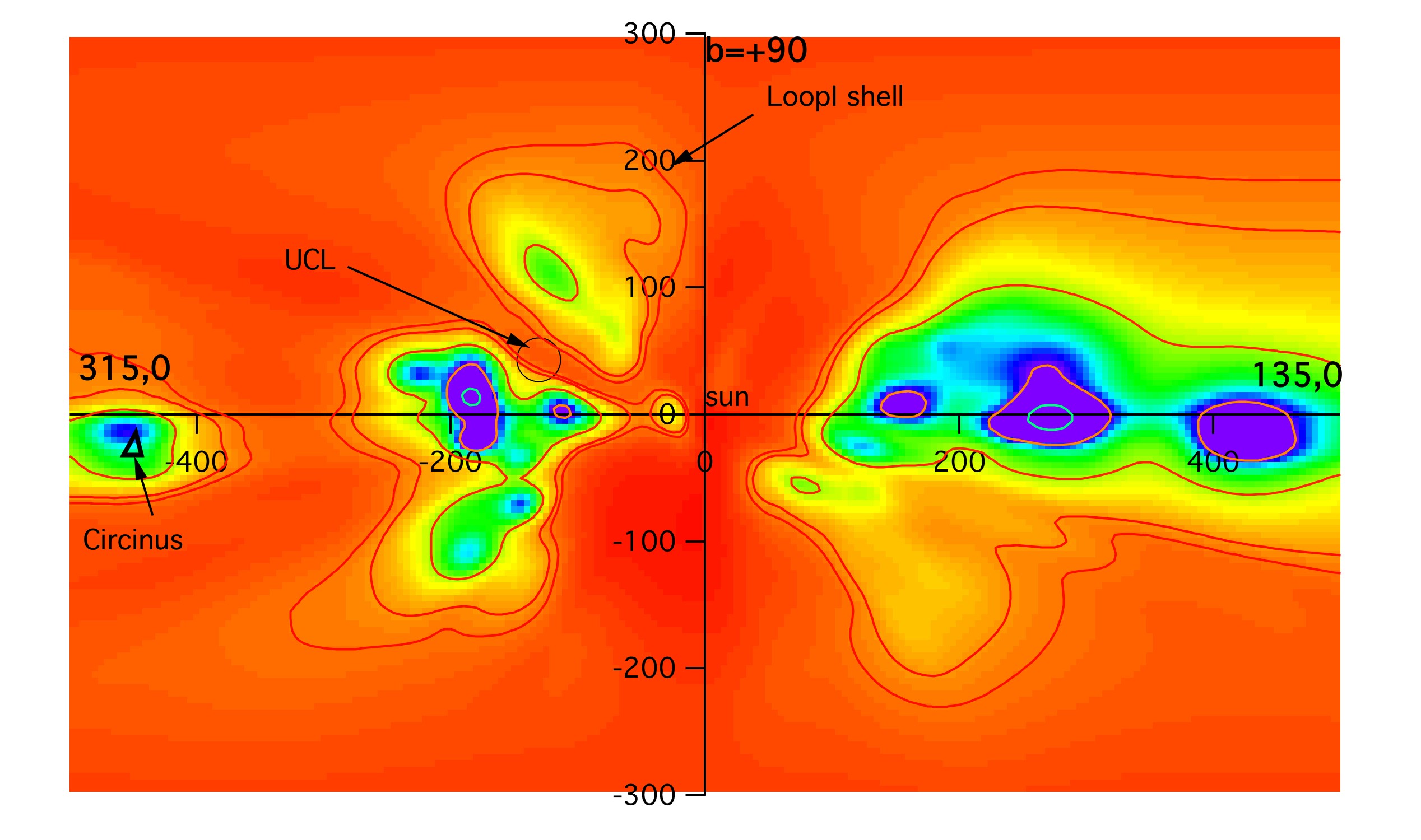}
	\includegraphics[width=\linewidth]{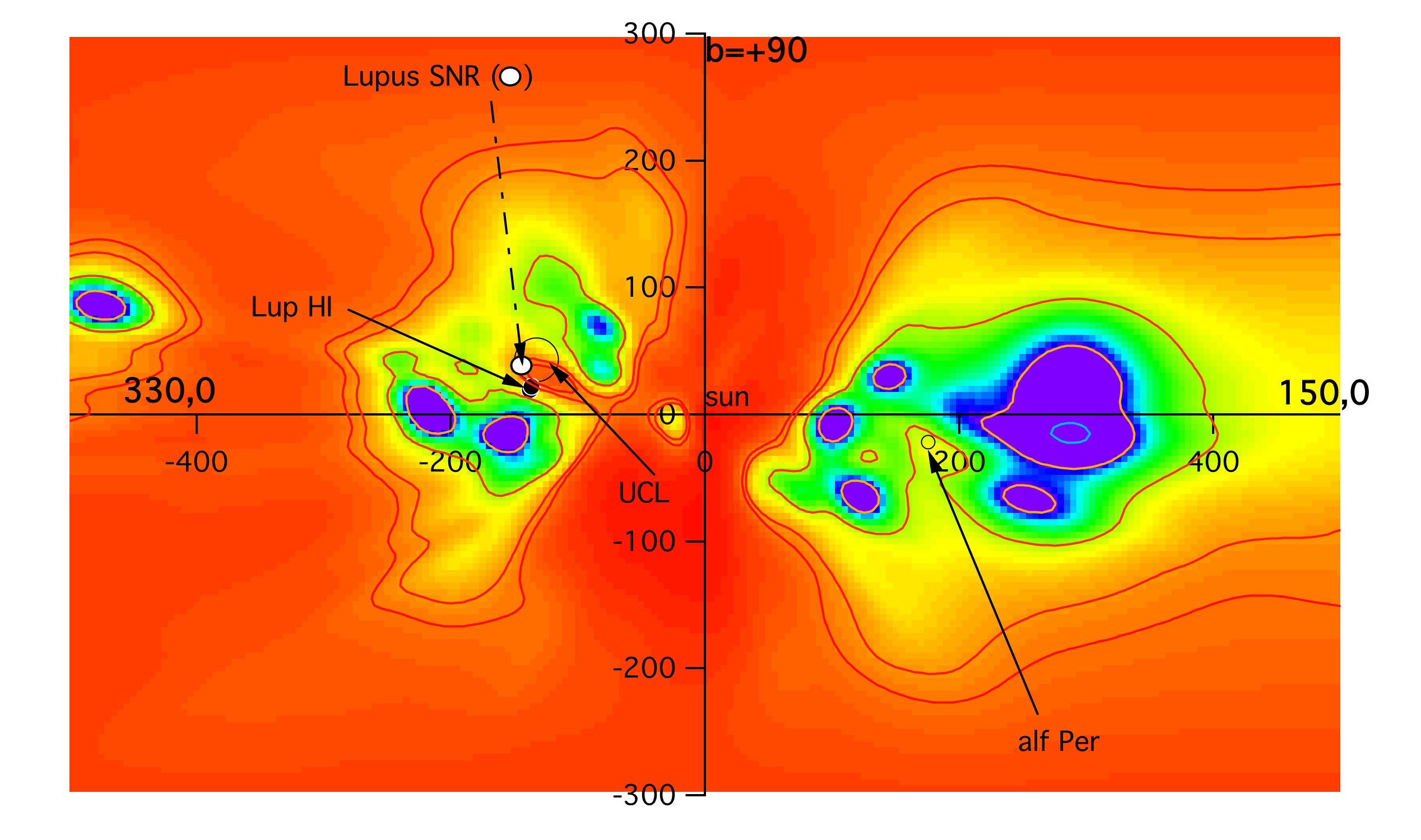}
	\includegraphics[width=\linewidth]{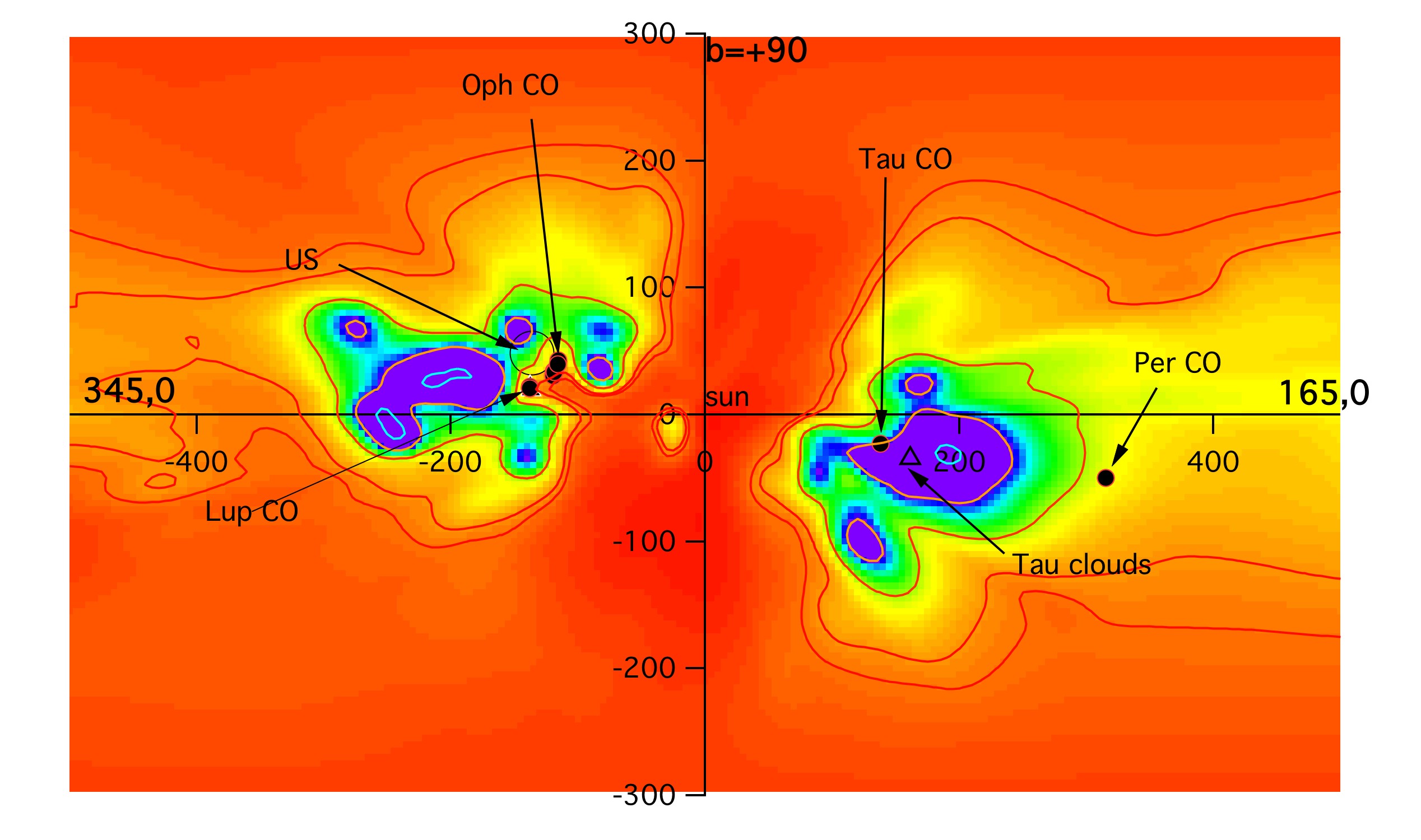}
\end{minipage}\hfill
\caption{Same as fig \ref{fourvertplanes1} for the next three vertical planes.}
\label{fourvertplanes4}
\end{figure*}

\begin{figure*}
\centering
 \includegraphics[width=\linewidth,height=10cm]{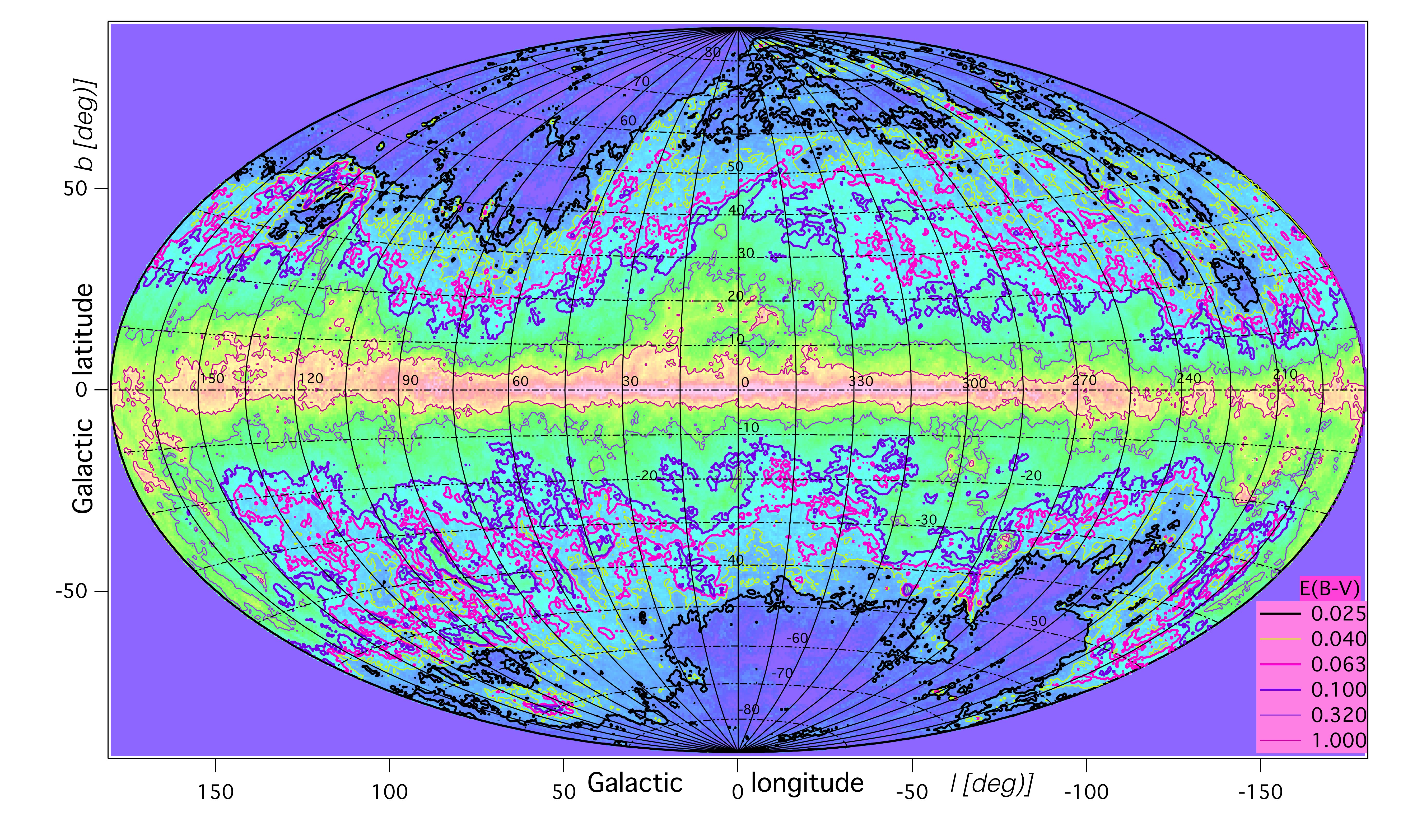}
 \caption{The Schlegel et al (1998) dust map is reproduced, along with the traces of the vertical planes of Fig \ref{fourvertplanes1} to \ref{fourvertplanes4} superimposed. The color-coded quantity is the logarithm of E(B-V). The six contours are for log(E(B-V)=0 (thin pink), -0.5 (thin violet), -1 (thick violet), -1.2 (thick pink), -1.4 (yellow) and -1.6 (black). Counterparts to the tenuous clouds at mid or high latitudes that appear in this SFD map can be searched for in Fig \ref{fourvertplanes1} to \ref{fourvertplanes4}, using the \textit{l,b} grid.}
  \label{fsd}%
 \end{figure*}

\begin{figure*}
   \centering
   \includegraphics[width=14cm]{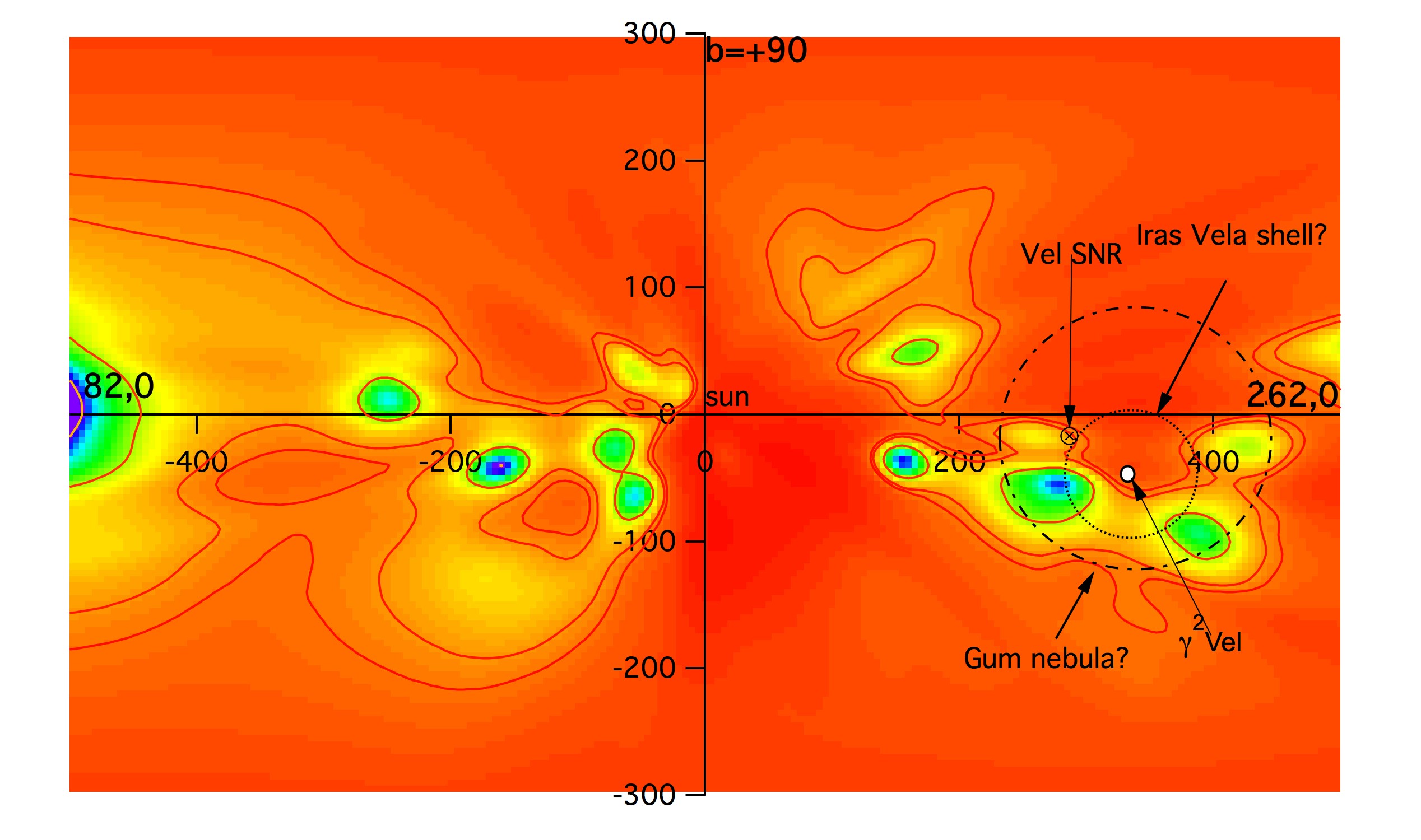}
   \caption{Same as fig. \ref{fourvertplanes1} for the \textit{l}=82-262$^{\circ}$ plane. The Vela SNR and the Wolf-Rayet system $\gamma_{2}$Vel are shown. The locations of the Iras Vela shell (IVS) and Gum Nebula contours are drawn according to the scenario of \cite{sushch11}.}
              \label{figgum}%
    \end{figure*}

 \begin{figure*}
   \centering
   \includegraphics[width=14cm]{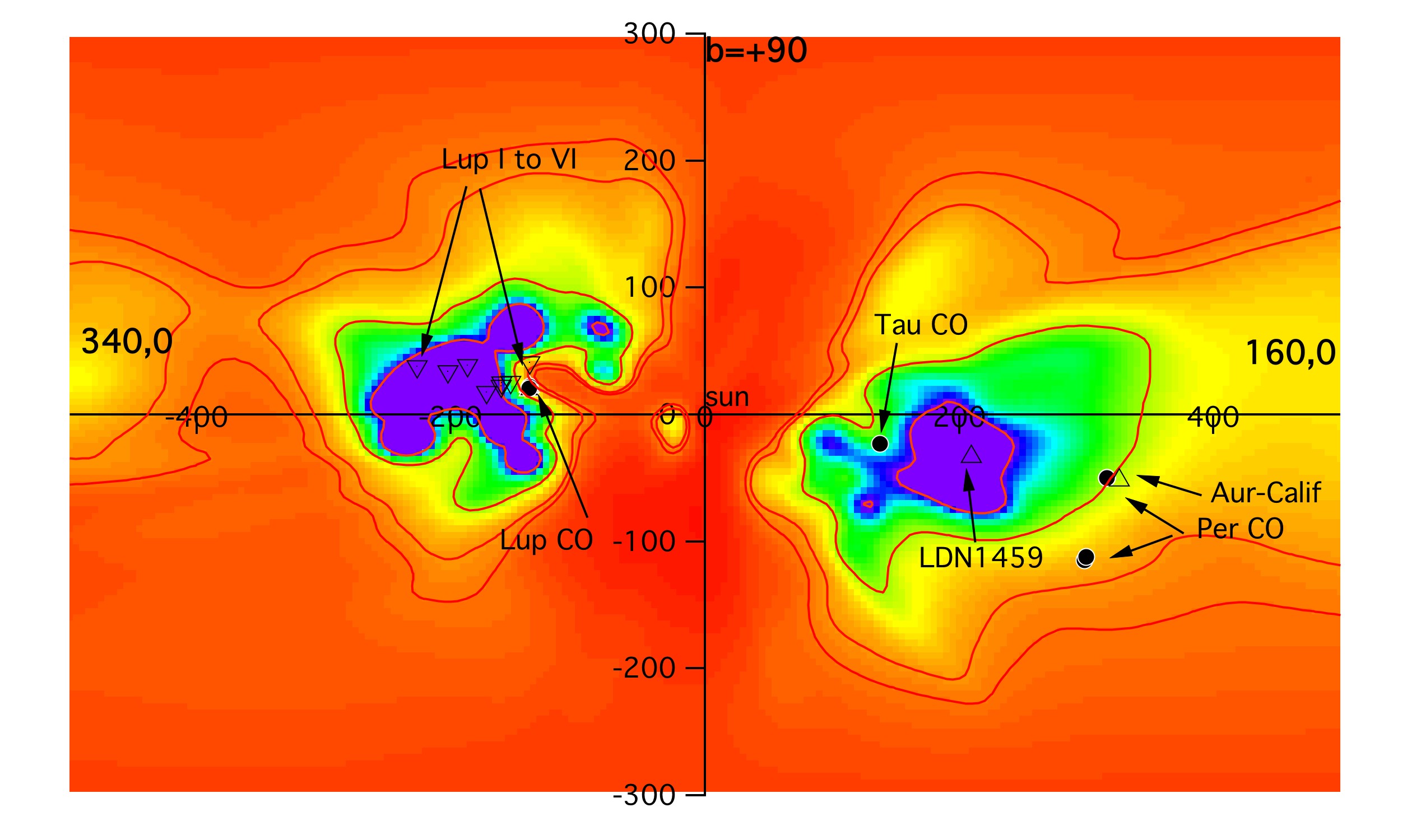}
   \caption{Same as fig. \ref{fourvertplanes1} for the \textit{l}=160-340$^{\circ}$ plane. Here are shown the locations of the Lupus molecular clouds derived by \cite{knude10}.}
              \label{vert160lupus}%
    \end{figure*}

 % Fig: maps for  15 deg spaced planes
%OB associations De Zeeuw 
%OK except Col 121 , but: Kaltcheva refutes the distace we also find dispersed values with new hipparcos
%alos see burningham 2003 un groupe jeune a 1050 reste est dispersé 
%aucune poussiere vu sur dobashi et schlegel 
%donc on confirme rien d'associe a col 121 a 500 pc 
%aussi groupe a 300 pas vu en poussieres
%seule chose vu vers 225 (dobashi e.g.) que l'on voit aussi plus prs

%orion cut \textit{l}=211
%on retourve le nuage CO centre a -15 alors que maddalena86 montre orion 500pc a b= -19 et monr2 a 213-13  830pc ou +
%donc on est un peu au dessus

%The distribution of the targets is far from regularly distributed in 3D. There are regions with high concentration of target stars, while in other regions are deficient.

 \begin{figure*}
   \centering
   \includegraphics[width=10cm]{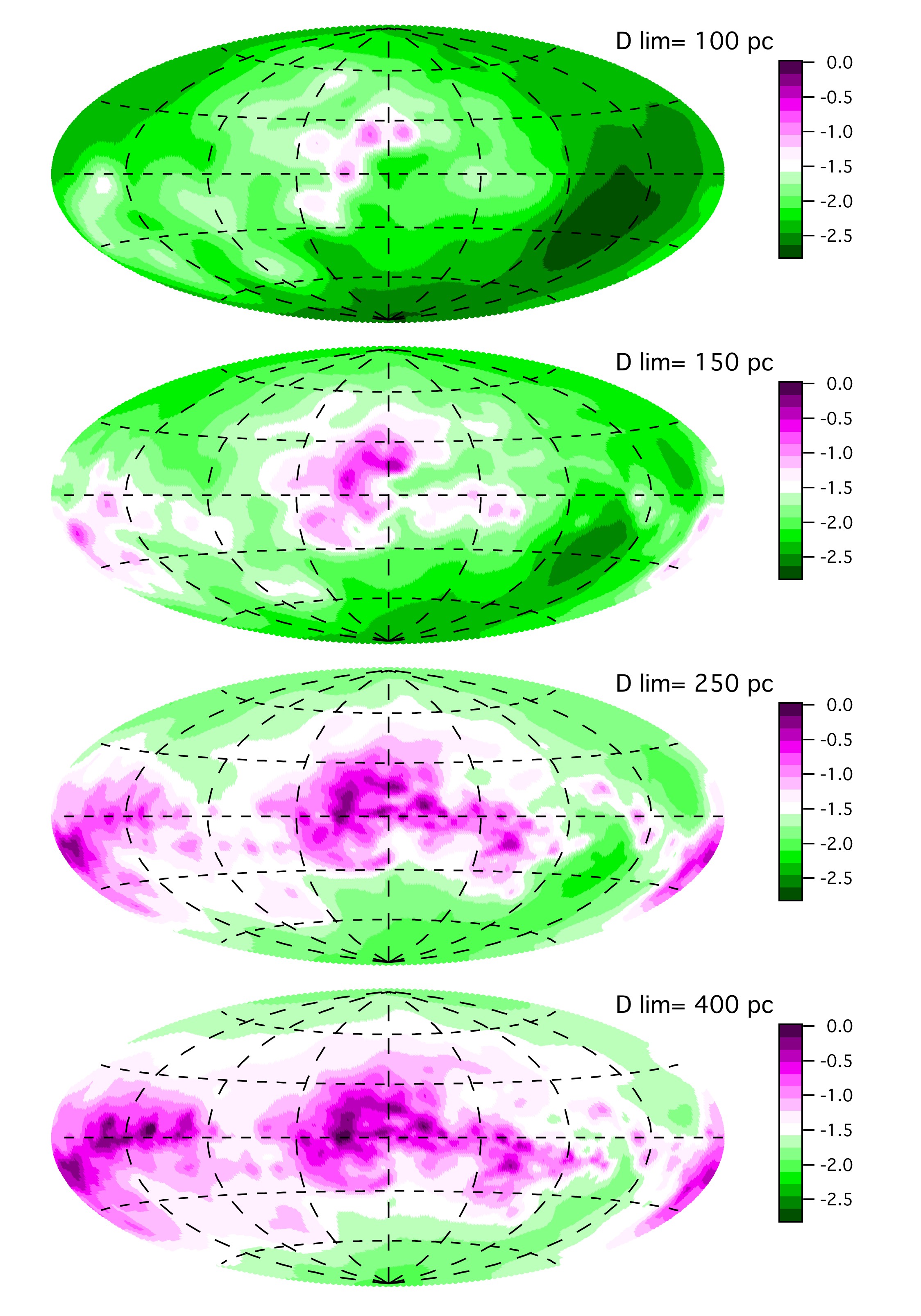}
   \caption{Opacity E(B-V) from the Sun to the distance D$_{lim}$=100, 150, 250 and 400 pc respectively (top to bottom), in logarithmic scale. The maps are in Aitoff projections, with meridian traces for \textit{l}=45, 90, 135, 225, 270 and 315$^{\circ}$ and parallel traces for \textit{b}=-60,-30, +30, and +60$^{\circ}$ shown as dashed and long-dashed lines resp. The color scale is identical in the four maps.}
              \label{2Dmaps1.pdf}%
    \end{figure*}

\begin{figure*}
   \centering
   \includegraphics[width=10cm]{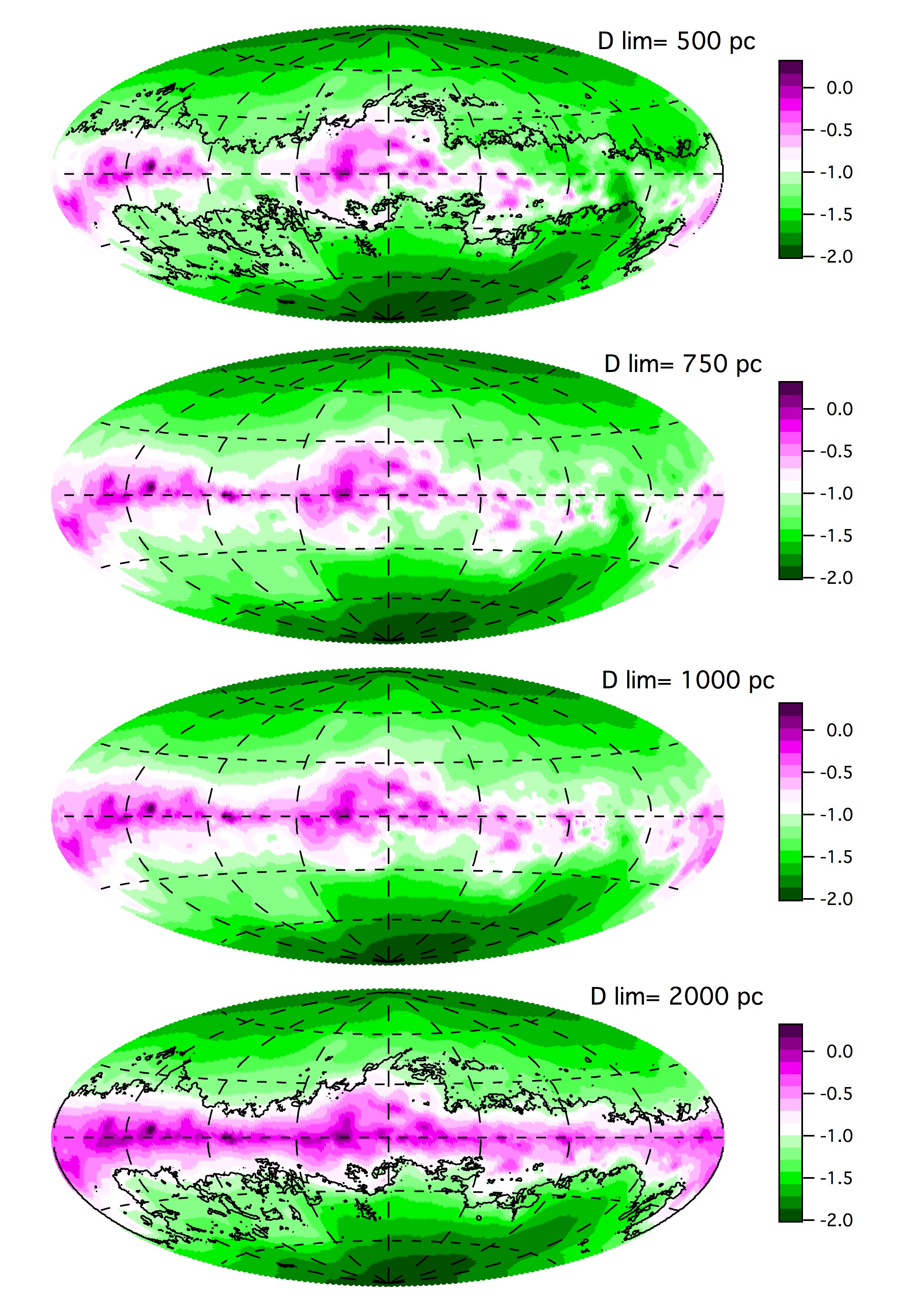}
   \caption{Same as fig \ref{2Dmaps1.pdf}, but for D$_{lim}$=500, 750, 1000, and 2000 pc and a different (and still unique) color scale. Superimposed on the first and last maps is the E(B-V)=0.1 iso-contour from the SFD map. E(B-V)=0.1 corresponds the white-to-green transition in the distance-limited opacity maps.}
              \label{2Dmaps2.pdf}%
    \end{figure*}

\begin{figure*}
   \centering
   \includegraphics[width=10cm]{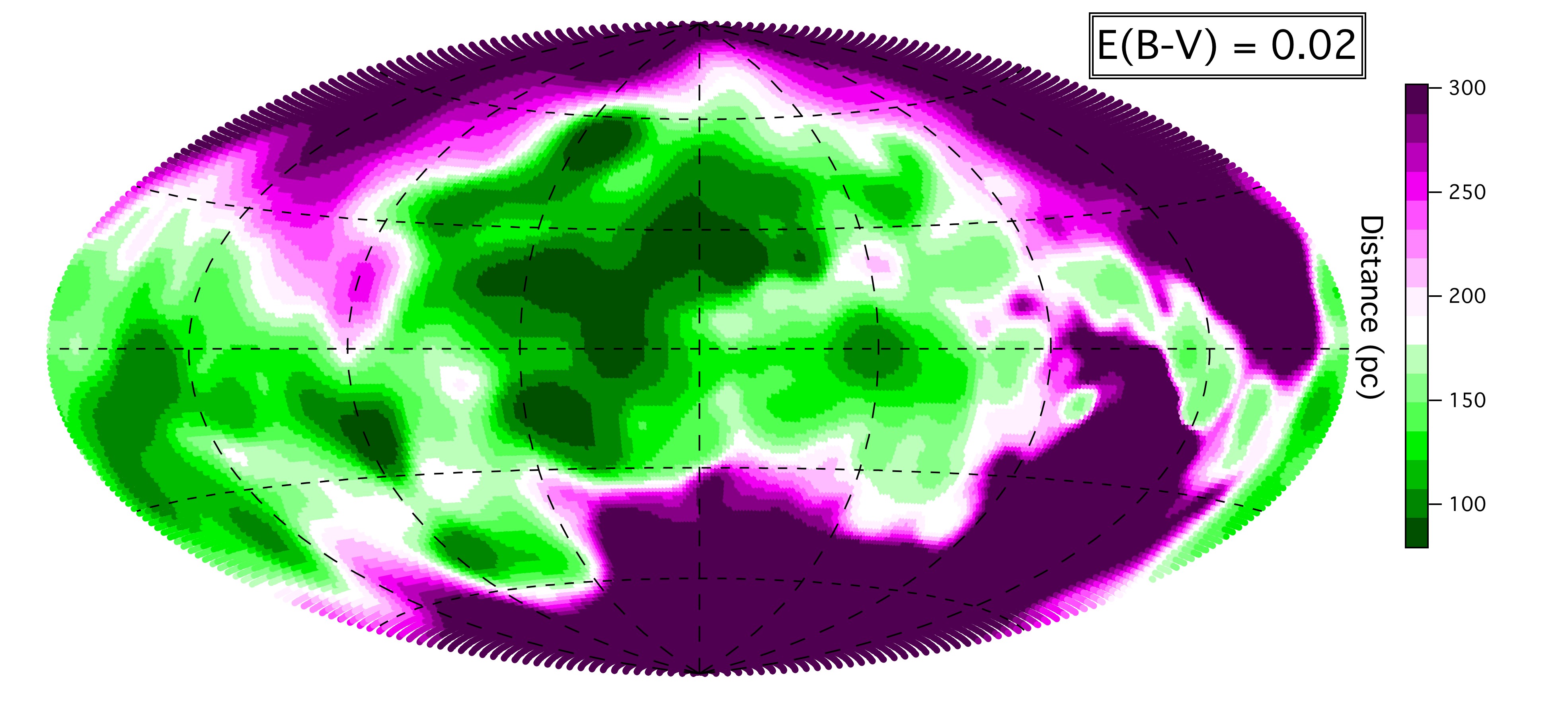}
   \caption{Distance from the Sun at which the integrated opacity, computed from the inverted differential opacity distribution, is reaching E(B-V)=0.02 mag. Directions corresponding to distances above 300 pc correspond to the purple color. The minimum value (darkest green) is $\simeq$ 80 pc.}
              \label{dist_ebv002}%
    \end{figure*}

\begin{figure*}
   \centering
   \includegraphics[width=10cm]{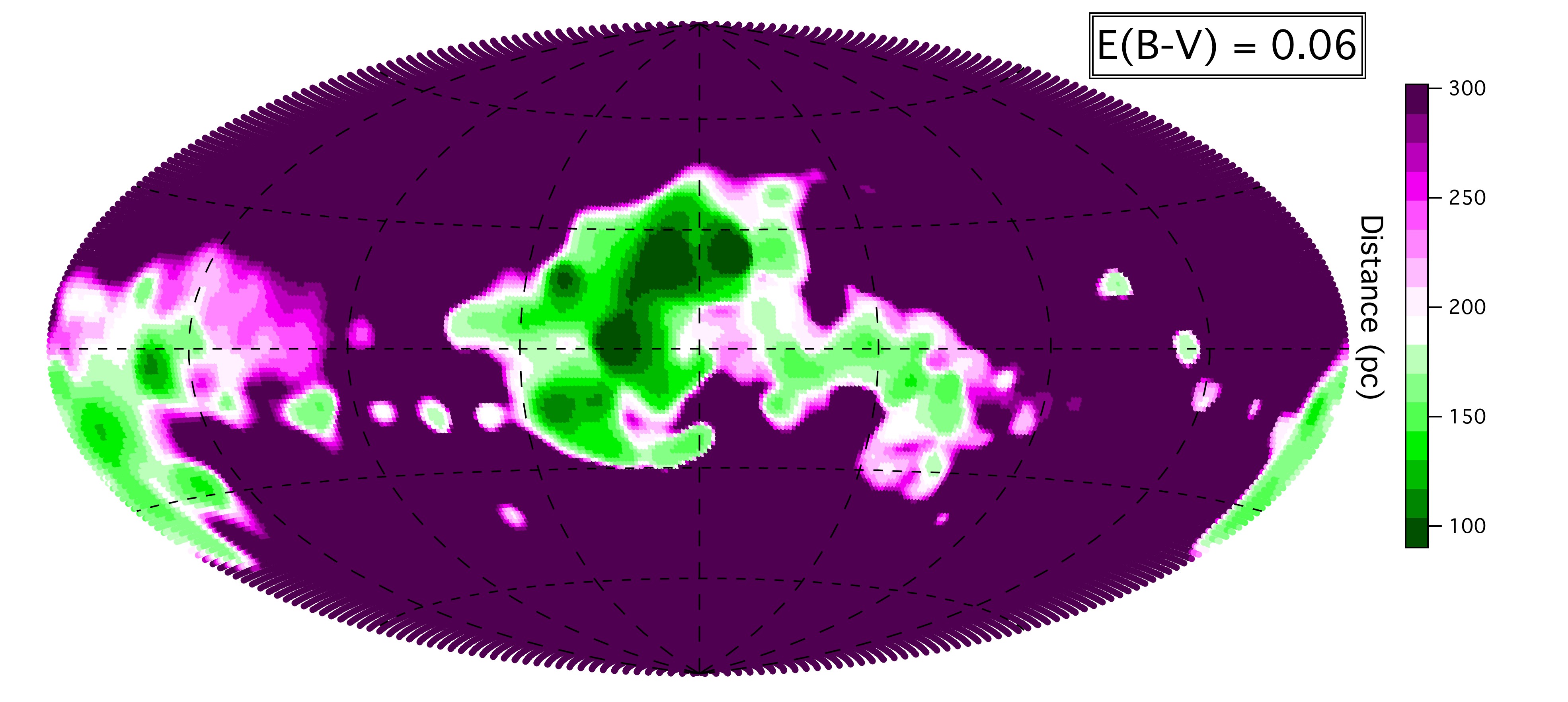}
   \caption{Same as Fig   \ref{dist_ebv002} for E(B-V)=0.06 mag. The minimum value (darkest green) for the distance is $\simeq$ 90 pc.}
              \label{dist_ebv006}%
    \end{figure*}

\subsection{Extinction in vertical planes}

We show in Fig \ref{fourvertplanes1} to \ref{fourvertplanes4} the opacity distributions in a series of 12 vertical planes containing the Sun, two consecutive planes being separated by 15$^{\circ}$ in longitude. For each plane we superimpose, when they happen to be located within 25 pc from it, the locations of the central parts of the nearby OB associations as they have been estimated by \cite{dezeeuw99}, as well as the HI and CO clouds listed by \cite{perrot03} and whose distances have been estimated by  \cite{degeus90, grenier89, madda86,murphy85,murphy86} and \cite{unger87}. This allows to check that those structures have their counterpart under the form of a dense region (for the clouds) or a cavity closely bounded by a dense region (for the OB associations) in the inverted maps. Note that we did not include Collinder 121 whose distance has been the object of several discussions (e.g. \cite{kaltcheva07, burningham03}) and maybe much larger than 500 pc. The new Hipparcos determinations of \cite{vanleeuwen07} indeed show that there are various distance ranges for the members. Interestingly our maps do not reveal any cloud at $\simeq$ 500 pc in the direction of Col121 (\textit{l,b}=236,-10). We also indicate the distances to the main nearby molecular clouds (or cloud complexes) derived by \cite{knude10} based on Hipparcos and 2MASS. 

One of the main advantages of the present method and of the dataset is the large number of nearby stars for which very low extinctions have been measured. This allows to locate nearby tenuous clouds.  Indeed, all vertical planes reveal low density structures above the Plane that are isolated or linked to denser structures, and whose shapes do appear quite complex. Because these structures are faint and the target star density decreases with the distance to the Plane, they appear blurred sometimes, but the comparison between their locations in the maps and the coordinates of the medium and high latitude clouds that show up in the 2D dust maps of  \cite{schleg98} (hereafter SFD) leaves no doubt about their reality.  To allow the reader to make those comparisons we have superimposed on the SFD dust map the traces of the vertical planes of Fig \ref{fourvertplanes1} to \ref{fourvertplanes4}. This is shown in Fig \ref{fsd}. We use a logarithmic scale for E(B-V) in order to identify more easily the faint medium and high latitude clouds. Interestingly, almost all features seen in Fig \ref{fsd} along those meridian lines have a counterpart under the form of an enhanced opacity in the corresponding vertical plane. 

As a first example we consider the \textit{l}=105-285$^{\circ}$ vertical plane (Fig \ref{fourvertplanes3} middle).

 i) The \textit{l}=105$^{\circ}$ half-plane North: in the SFD map (Fig \ref{fsd}) there is a Northern extension of the dust (yellow contour  up to  \textit{b}=+45$^{\circ}$) and many very small features (seen in yellow) at higher latitudes up to \textit{b}=+60;+80$^{\circ}$). In Fig  \ref{fourvertplanes3} (middle), just above the main concentrations extending from the Plane up to \textit{b}=+30;+40$^{\circ}$ (and colored violet, green, yellow), is a fainter feature apparently linked to the bulk but extending further up to +45$^{\circ}$ (faint yellow-orange) and located at $\simeq$200 pc. We believe this is the counterpart of the cloud seen in the 2D map. In addition, there is a faint dust cloud located much closer ($\leq$50 pc), that must produce small absorptions up to +80$^{\circ}$ latitude and probably explains the very faint, high latitude features in the 2D map. 

ii) The \textit{l}=285$^{\circ}$ half-plane North: in Fig \ref{fsd} it can be seen that compared to the previous half-plane the opacity is stronger at high latitude. More quantitatively, the E(B-V)=0.063 (thick pink) contour extends up to \textit{b}=+35$^{\circ}$, instead of +20$^{\circ}$. The E(B-V)=0.04 (yellow) contour extends up to +55$^{\circ}$, disappears  above, then a  more opaque area reappears at \textit{b}=+55$^{\circ}$. A look at Fig  \ref{fourvertplanes3} (middle) indeed shows nearby dust (green) up to \textit{b}$\simeq$+55$^{\circ}$ and faint extensions visible almost up to the pole.

 iii) \textit{l}=The 105$^{\circ}$ half-plane South: in Fig \ref{fsd} the E(B-V)=0.063 (thick pink) contour extends down to \textit{b}=-55$^{\circ}$, which corresponds to the cloud seen at $\simeq$170 pc in Fig \ref{fourvertplanes3} (pale green and yellow). On the other hand, the 80pc distant feature at lower latitude (\textit{b}=-65$^{\circ}$) in Fig \ref{fourvertplanes3}) may be the counterpart to the \textit{b}=-70$^{\circ}$ small cloud seen in Fig \ref{fsd}. 
 
 iv)The  \textit{l}=285$^{\circ}$ half-plane South: in Fig \ref{fsd} the E(B-V)=0.063 (thick pink) contour extends down to \textit{b}=-45;-50$^{\circ}$, which corresponds to the cloud seen at 150 pc in Fig \ref{fourvertplanes3} (dark blue, green, yellow). The -50;-60$^{\circ}$ small extension seen in Fig \ref{fsd} has no clear counterpart but may be simply the Southern extension of the previous feature (faint yellow). Alternatively  it may be more distant than 350pc, or missing due the lack of constraining targets.

Comparing in the same way all half-planes shows that most of the high latitude features in Fig \ref{fsd}  are found in the maps. As a second example we consider the \textit{l}=15-195$^{\circ}$ vertical plane in Fig \ref{fourvertplanes1} (middle). At \textit{l}=15$^{\circ}$, the inverted map shows clearly in the orth  dense high latitude clouds up to \textit{b}=+60$^{\circ}$, with a small extension at even higher latitude (around +70$^{\circ}$) that is seen up to 150 pc (in the same direction there is also a very faint  cloud much closer to the Sun). This is similarly found in Fig \ref{fsd} along the Northern part of the \textit{l}=+15 meridian. In the South, Fig \ref{fourvertplanes1} (middle) reveals a dense cloud at \textit{b}=$\simeq$-25;-30$^{\circ}$ and at d$\simeq$150 pc, whose emission is clearly marked in the Fig \ref{fsd} (green color). At \textit{l}=195$^{\circ}$ there is a strong contrast between the Northern latitudes, where an appreciable opacity is reached as low as \textit{b}$\lesssim$+20$^{\circ}$  (transition between yellow and orange), and the South where dust extends down to \textit{b}=-55;-60$^{\circ}$. The \textit{l}=195$^{\circ}$ meridian in Fig \ref{fsd} indeed shows clearly this low latitude cloud extension (pink contour) while the same level of dust emission does not go beyond +20$^{\circ}$ in the North. 
We note however a discrepancy between the total absence of opacities detected in the North in the inverted maps above \textit{b}$\simeq$+30$^{\circ}$ (at this latitude there is a small feature at +450 pc) and the SFD map that shows E(B-V)=0.04 up to +42$^{\circ}$. This discrepancy seems to exist for the whole interval \textit{l}=180-230 $^{\circ}$. This dust may be much more distant than 1kpc, or missing in the maps due the lack of constraining targets.
Apart from this region agreements exist for all directions, and it is possible to assign a distance to the Fig \ref{fsd} high latitude emission features by using the appropriate maps. This suggests that our limitation of Z=300 pc for the target stars had no significant influence on the results.

% As we have seen, all high-latitude clouds that are visible in the SFD map for E(B-V) above $\simeq$ 0.025-0.040 (black and yellow contours) are found to have counterparts in the vertical maps. %There is one exception in the North at longitudes 180-230 $^{\circ}$, where there are no or very small counterparts to the high latitude features in the SFD map. This is possibly due to the scarcity of target stars beyond about 100 pc in those directions, and deserves further studies. However we note that in the Planck maps do not show the same level 

Given the limited number of faint stars the mapping is in some case less performant for the very dense structures, in the sense that dense clouds may be underestimated due to the lack of strongly reddened targets, or their densest areas appearing slightly displaced with respect to the actual cloud centers. This is e.g. the case for the Orion structures  in the \textit{l}=210$^{\circ}$ half plane (Fig \ref{fourvertplanes1} bottom) that are centered at \textit{b}=-15$^{\circ}$ while the CO clouds are located a few degrees below. Also, their extent is not as wide as in the 2D maps and they do not appear as opaque as the dust maps predict. Such biases due to the target star distribution and the inversion should disappear by increasing the extinction databases.

Several planes are particularly interesting since they contain well-studied regions and in some case the maps shed some light on their distances. E.g. the \textit{l}=30$^{\circ}$ half plane crosses the Aquila rift (Fig \ref{fourvertplanes1}, bottom). We find a cloud complex at low latitude (\textit{b}=5$^{\circ}$) that starts at $\simeq$190 pc and centered at d$\simeq$220 pc. It corresponds to the Serpens cloud  and its recent distance assignment by \cite{knude11} (193 pc), and \cite{strai03} (225 pc). We note that there is a second cloud complex at $\simeq$420 pc that appears at lower latitude in our map. Its distance  is in perfect agreement with the 420 pc distance determination of the star EC95 by \cite{dzib10}, also believed to be part of the Serpens cloud. We consider the alignment of the two different clouds as a potential explanation for the discrepancy between the two measurements: the 420 pc cloud may extend at slightly higher latitudes than in our inverted maps, its Northern part being not part of the maps due to the absence of targets stars located beyond the Serpens cloud (those stars are too much extincted). EC95 could be part of it and seen in projection against the Serpens clouds.

There is another potentially interesting feature, this time in the opposite half-plane \textit{l}=210$^{\circ}$. We have drawn in Fig \ref{fourvertplanes1} (bottom)  the two directions that mark the limits of  the Northern and Southern parts of the Barnard's Loop crescent. Interestingly they correspond to two elongated parts of a cloud located at $\simeq$170 pc, while the Loop is believed to be associated to the extended Orion region at 440 pc, also visible in the map beyond the 170 pc cloud. We believe that this coincidence deserves further studies of the distance to the Barnard's Loop. 

We have also displayed in Fig \ref{figgum} the \textit{l}=82-262$^{\circ}$ plane. This specific plane is particular interesting due to the presence of the several objects that gave rise to many discussions and analyses: the Vela SNR, the Wolf-Rayet system $\gamma_{2}$Vel, the Iras Vela shell (IVS), and the Gum Nebula. In particular, the four were modeled as part of a global scenario by \cite{sushch11}. The locations of the clouds derived by inversion agree well with this recently devised scenario.

Similarly the \textit{l}=160-340$^{\circ}$ plane drawn in Fig \ref{vert160lupus} is particularly rich in clouds. We note here very good agreements between the maps and the cloud distances derived by \cite{knude10} for the series of Lupus clouds and LDN1459. The more distant Aur-California dark cloud does not correspond to a very dense region in the map, however this may be due to the screening effect of Taurus clouds and the lack of  targets beyond them.

\section{Distance-limited 2D reddening maps}

In order to facilitate the comparison of the inverted 3D distribution with 2D  maps based on emission data, we have computed the integral of the differential opacity up to various, selected distances D$_{lim}$ and for one degree steps in longitude and latitude to cover the whole sky. The resulting maps are shown in Fig \ref{2Dmaps1.pdf} and \ref{2Dmaps2.pdf}. The series of maps at increasing distances from D$_{lim}$=100 pc to D$_{lim}$=2 kpc allows to infer at which distances the most conspicuous clouds start to get visible in the 2D maps. On the first and last map of Fig \ref{2Dmaps2.pdf}, computed for a radius of 500 pc and 2 kpc resp. we have superimposed the E(B-V)=0.1 iso-contour of the  SFD map. It can be compared with the log(E(B-V))=-1 iso-contour in the 500 pc and 2 kpc integrated maps (not drawn but easily seen thanks to the white-green transition). 
Although  the inverted maps are lacking details, the similarities between the two E(B-V)=0.1 contours in the  2 kpc map show that in all directions that define this contour most of the extinction measured by SFD  is actually generated closer than 2 kpc and retrieved through the inversion. At variance with the D$_{lim}$=2 kpc map, in the D$_{lim}$=500 pc map the two contours are significantly different, especially at 90 and 270$^{\circ}$. In those directions the extinction is generated beyond 500 pc. 

It is important to note that many of the weak features at mid- and high latitudes that appear in the vertical planes in Fig \ref{fourvertplanes1} to  \ref{fourvertplanes4} are not seen in the integrated extinction maps for D$_{lim}$ beyond $\simeq$200-300 pc. This is mainly due to the high differential opacity value chosen for the prior distribution. As a matter of fact, we used a large height scale h$_{0}$=200 pc, which results in a small decrease of the differential opacity with increasing distance from the Plane. This high prior opacity  makes the integrated opacity of the tenuous clouds weak or negligible in comparison with the full integrals. However, as we explained above, in practice this large scale height  facilitates the detection of the nearby weak features. A typical example of this effect is the case of the North Celestial Pole Loop. This arch that extends from \textit{l}$\simeq$120$^{\circ}$ to \textit{l}$\simeq$170$^{\circ}$ and culminates at \textit{b}$\simeq$=40$^{\circ}$ is clearly seen in the SFD or Planck maps. It appears as a broad extension at high latitude in the D$_{lim}$=250 pc map (it is located at $\simeq$ 200 pc) but does not appear in any of the 2D maps for D$_{lim}$$\geq$ 500 pc. As a conclusion, the search for the distances to the high latitude thin clouds should be performed essentially in the planar cuts in the 3D distribution.

We show in Fig \ref{dist_ebv002} (resp. Fig \ref{dist_ebv006}) 2D maps of the distance at which the integrated opacity, computed from the inverted differential opacity distribution, is reaching the limiting value E(B-V)=0.02 (resp. 0.06) mag. It allows to visualize where the Local Cavity boundaries are the closest. The closest part of the Aquila Rift and the upper Scorpius and Centaurus regions appear conspicuously.  The comparison of those maps as well as Fig \ref{2Dmaps1.pdf} and \ref{2Dmaps2.pdf} with the results in distance bins of \cite{reis11} show many similarities.  More maps of this kind will be presented in a future paper analyzing soft X-ray background data in conjunction with the 3D maps.  

\section{{3D visualization}}

{The last figure of this article is interactive and represents an iso-differential opacity surface corresponding to 0.0004 mag. pc$^{-1}$, made with the YT software (\cite{yt_ref}). It allows to visualize the global distribution of the main cloud complexes.  Such a representation on the other hand does not allow to visualize details  within those surfaces, nor to visualize cavity contours. Two additional interactive 3D images allowing transparency and showing more structures can be seen at: } \\
\url{http://mygepi.obspm.fr/~rlallement/ism3d.html}    \\
{and } \\
\url{http://mygepi.obspm.fr/~rlallement/ism3dcrevace.html} \\
%mygepi.obspm.fr/\~{}rlallement/ism3dsurf1.html    \\
{Note however that faint features must be searched for in the slices presented above.}

%\begin{figure*}
%\centering
%\includemovie[
%	poster,
%	toolbar, %same as `controls'
%	label=surface2.u3d,
%	text=(surface2.u3d),
%	3Daac=0.000000, 3Droll=0.000000, 3Dc2c=-13.010000 2431.580078 -153.699997, 3Droo=2436.467529, 3Dcoo=13.007996 -27.415401 -153.699310,
%	3Dlights=CAD,
%]{\linewidth}{\linewidth}{surface2.u3d}
%\end{figure*}

%\begin{figure*}
%\centering
%\includemovie[
%	poster,
%	toolbar, %same as `controls'
%	label=surface2000rho4.u3d,
%	text=(surface2000rho4.u3d),
%	3Daac=60.000000, 3Droll=0.000000, 3Dc2c=-0.000000 5689.000000 -0.000000, 3Droo=5689.000000, 3Dcoo=-0.000000 0.000000 0.000000,
%        3Dlights=CAD, 3Drender=Solid, 3Dbg=0 0.8 1
%]{\linewidth}{\linewidth}{surface2000rho4.u3d}
%\label{surface2000rho4.u3d}
%\end{figure*}

\begin{figure*}
\centering
\includemovie[
	poster,
	toolbar, %same as `controls'
	label=biggerbiggercube.u3d,
	text=(biggerbiggercube.u3d),
	3Daac=10.000000, 3Droll=0.000000, 3Dc2c=0.000000 5689.000000 0.000000, 3Droo=5689.000000, 3Dcoo=-0.000000 0.000000 0.000000,
	3Dlights=CAD,
]{\linewidth}{\linewidth}{biggerbiggercube.u3d}
\label{surface2000rho4.u3d}
\caption{Interactive 3D iso-opacity surface for 0.0004 mag. pc$^{-1}$. The Sun is represented by a cube.}
\end{figure*}

%__________________________________________________________________
\section{Discussion}

We have presented new differential opacity maps of the local ISM based on the merging of several photometric catalogs and associated color excess measurements. The data are inverted using a Bayesian code based on the \cite{tarantola82} theoretical work and developed by \cite{vergely01} and \cite{vergely10}. New spatial correlation kernels have been introduced in order to reproduce the ISM structure in a better way, and the quantitative parameters have been chosen in a very conservative way to avoid non realistic condensations. The maps extend the earlier results of \cite{vergely10} to distances of 800-1000 pc at low latitudes and 300 pc below and above the Plane. 

Our main findings are the following:  

-The comparison with the former maps shows a good agreement despite the use of a different reddening database and different correlation kernels. In particular, the addition of a large number of early-type stars did not impact on the local cavity boundary mapping. The main clouds are located in 3D space in a way that is coherent with other independent determinations, and with 2D dust maps. This demonstrates that the present inversion method gives satisfying results provided parameters are carefully and conservatively chosen in accordance with the number of sightlines entered in the inversion process. 
%As a matter of fact there is good  agreement with the previous (smaller) maps, found despite the use of different kernels,  and agreement on the location of the main OB associations and related clouds with other estimates. As a consequence it will  be appropriate for futures inversions of larger datasets, and hopefully to data from the Gaia mission, when available. 

-The combination of the inversion technique and the dataset characteristics is particularly appropriate in revealing nearby interstellar cavities. The geometry and distribution in sizes of those various cavities is available for comparisons with three-dimensional hydrodynamical models of the ISM and bubble evolutions under the repetitive action of stellar winds and supernovae (\cite{avillez09}).  In particular the maps reveal a huge cavity we identify as the super-bubble GSH238+00+09 (\cite{heiles98}) and an elongated cavity in the opposite direction. By giving a global perspective on the distribution of the main clouds and cavities, the new maps may also help shedding light on the formation of the Gould belt/Lindblad ring structure, e.g. the scenario devised by \cite{olano01} that attributes the formation of the local arm and the Gould belt to a supercloud having entered into a major spiral arm 100 Myr ago (see also the recent work of \cite{perrot03}). Indeed, the majority of the cloud complexes seem to surround the chain of cavities formed by  GSH238+00+09, the Local Bubble and the \textit{l}=70$^{\circ}$ cavity, and this clearly points to a special role of this 60(70)$^{\circ}$-240(250)$^{\circ}$  direction in the local ISM history. We note that this direction is the one of the interstellar helium ionization gradient axis found by \cite{wolff99}, a potential additional consequence of  the event that gave rise to the whole structure.

-The inversion technique and the dataset are also particular appropriate to locate the faint structures that lie above or below the Plane (E(B-V)$\leq0.1$), and are apparent in the emission maps (\cite{schleg98}). Such faint and angularly extended clouds cannot be mapped using statistical techniques. The majority of the latter features are located in 3D and found to be close, at less than $\simeq$ 150 pc. While there is definitely an inclined cavity fully devoid of dust (linking the local {\it chimneys} to the halo, around this cavity the dust pattern is quite complex. Most faint clouds like the Northern Loop1 arches and the North Celestial Pole Loop are located between 100 and 200 pc. 

-The maps can be used to potentially disentangle coincidental similarities in directions for unrelated clouds at different distances. We have discussed two particular cases.

We must caution that, in addition to the limitations in spatial resolution, that are linked to the use of a smoothing length, the present dataset is not appropriate for a detailed mapping of the dense clouds. As a matter of fact there is a limitation in the brightness of the target stars, and the subsequent lack of strongly reddened stars results. This results in a poor representation of regions beyond opaque clouds,  badly sampled. There is for the same reasons a bias towards low opacities and some of the opaque clouds are under-represented. Hopefully in future additional data towards weaker targets will allow to eliminate those biases and the decrease of  the mean distance between targets will allow to use shorter smoothing lengths, leading to a better spatial resolution. This is especially mandatory for the dense atomic and molecular phase. This should help understanding the complex kinematical structure of the whole Sco-Cen area recently derived by \cite{poppel10} based on HI 21 cm data from the Leiden-Dwingeloo Survey, and its link with the Gould belt formation. Present or future surveys and especially the Gaia mission should help improving the mapping. The use of the kinematics that is available from gas emission or absorption data should also help disentangling the structures and lead to better maps. 
%A web-based and open tool allowing to navigate in the three-dimensional distribution and download maps is devoted to future work.
% Distributions of the differential opacity in specific planes that are not parts from this article can be obtained on request from the authors. 

%xxxIn a future paper (Puspitarini et al.), we provide a first analysis of the resulting extinction maps by comparing them with soft X-ray background data.

\begin{acknowledgements}
J.L.V, R.L., and L.P.  acknowledge funding by the French Research Agency in the frame of the STILISM project.  We thank I. Grenier and J.M. Casandjian for useful discussions.  
\end{acknowledgements}

%%-----------------------------------------------------

%%-----------------------------------------------------

\newpage
\begin{appendix}

\section{Convergence and treatment of oscillations}

For most of the choices of parameters, the inversion takes about 3-4 iteration steps for convergence. We have checked that allowing for more steps does not modify significantly the distribution. The results here are obtained after 10 iterations. A number of values for the four main parameters, $\xi_0, \xi_1, \sigma_0, \sigma_1$, have been tested. We show here the results for one favored set of parameters. It corresponds to $\xi_0$= 30 pc,  $\sigma_0=0.6$, $\xi_1$= 15 pc and  $\sigma_1$=0.8. Our criteria for this choice of parameters are guided by a balance between the quality of the adjustments to the data and a conservative choice of the smoothing lengths. We are also helped by the distribution pattern itself, and in particular the appearance of elongated radial structures where the number of target stars is not sufficient with regard to the used kernel. However, we cannot avoid such radial structures at large distances where targets are missing if we want to keep a kernel appropriate to the nearby regions and uncover the nearby structures. This is illustrated in the Appendix.

Apart from the introduction of new kernels, the updated code includes now a new stage of convergence control.  Indeed, in some areas (the problem arose around \textit{l}=80$^{\circ}$ and b=0$^{\circ}$, a direction that corresponds to strong changes in the cloud properties with increasing longitude) the correlation length is large compared to the distances, thus the model has difficulties by fitting strong extinctions which are mixed with weak extinctions at distances much lower than the correlation length. Under these conditions, the model oscillates from one iteration to the other between the presence of a strong structure of extinction and the absence of extinction: two solutions are then possible for the model. In order to avoid these oscillations, at the end of the third iteration, one detects the extinction data which are further than three sigma from the model. In fact, the model practically converged in the majority of the areas of the Galaxy, except in the significant areas where the gradients in extinction are particularly important. The data which are further from the model show the areas where the model cannot reach the data. In this case, one increases the variance of the data by a factor of 2 ; this allows putting less weight at the data which deviate too much from the model. One starts again this filtering with each new iteration. From a practical point of view, about 300 sightlines are thus filtered in third iteration, from the total of 23,000. This number decreases from one iteration to the other and vanishes at the end of the sixth iteration. After convergence, the estimated model privileges smoothed solutions with low extinction values. 

\section{Distribution at large distance: constraint limitations}

\begin{figure*}[h!]
   \centering
   \includegraphics[width=10cm]{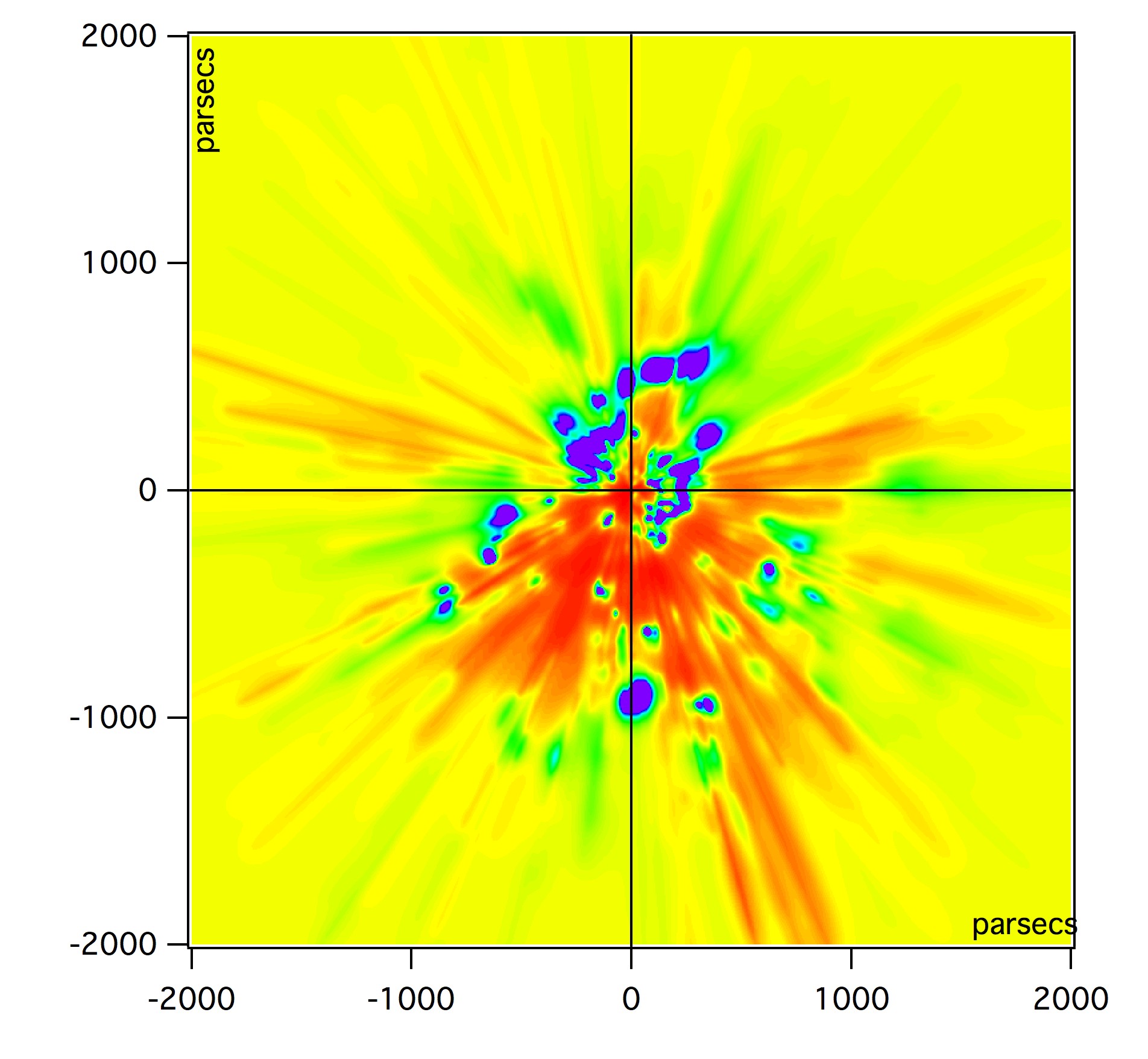}
   \caption{Galactic Plane cut in the 4000 pc 3D distribution of inverted opacity. This map clearly shows the limits between the prior distribution (homogeneous color) and areas where the model has been constrained by the dataset. The elongated radial structures correspond to directions where cavities over dense areas are detected but the number of target stars is too small to constrain their location, hence the spreading over a large distance.}
              \label{FigfullGP}%
    \end{figure*}

While the inversion has been performed in a 4000$\times$4000$\times$600 pc$^{3}$, the scarcity of distant stars results in very low or null constraints beyond $\simeq$800 pc. In this case the prior distribution is kept unchanged. We show in Fig \ref{FigfullGP}  the computed inverted distribution in the galactic Plane up to + or -2kpc,  in order to illustrate the limits of the area that is actually constrained. 

\section{Target distribution}

\begin{figure*}[h!]
   \centering
   \includegraphics[width=12cm]{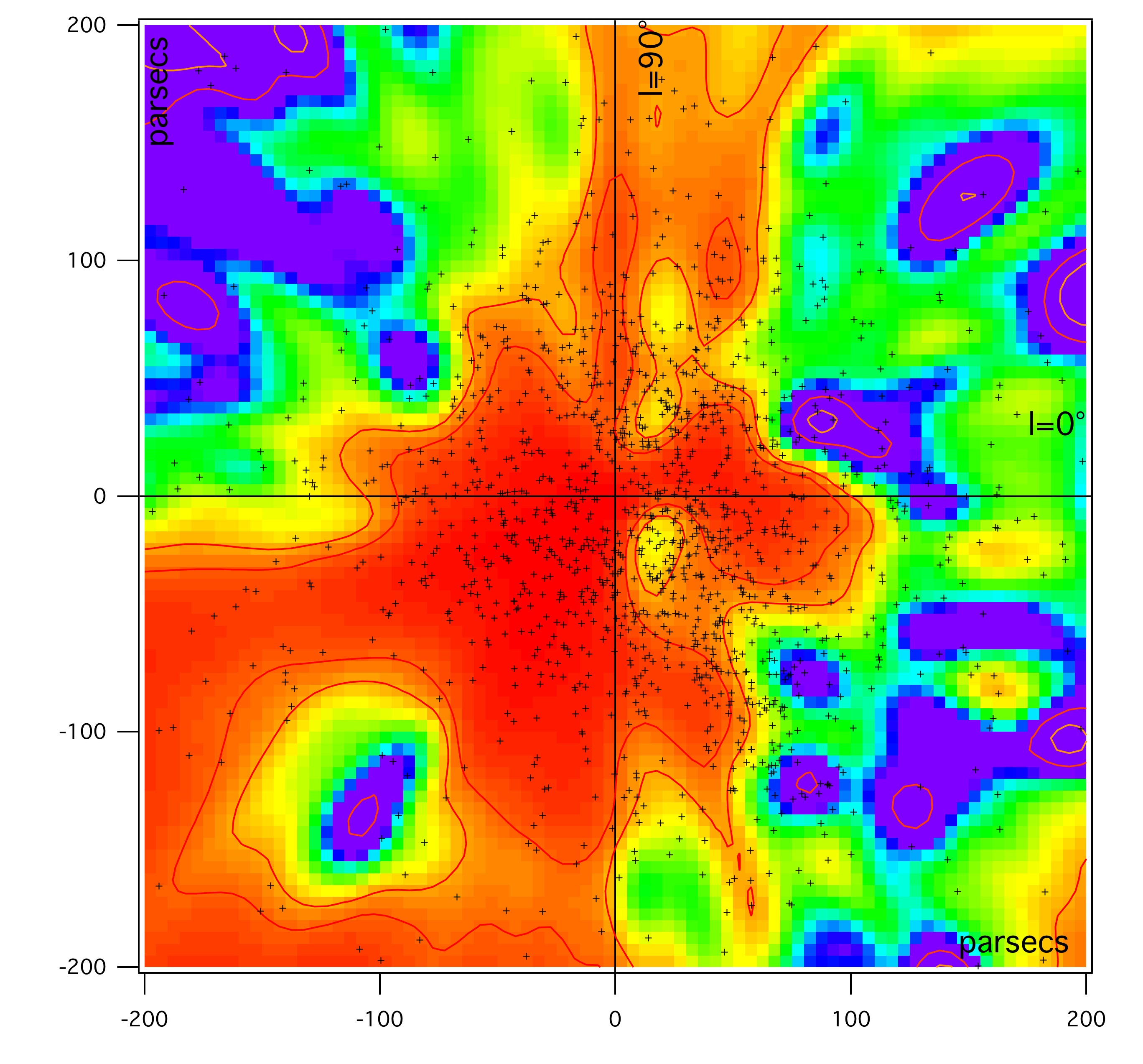}
   \caption{Galactic Plane cut in the 3D distribution of inverted opacity with stars within 10 pc from the Plane superimposed (plus signs). The distribution  shows the regions that are well constrained by the data, and how, given the present limited dataset, nearby cavities are better represented compared to cloud complexes. The distribution also allows to figure out the limiting size for the structures from the distance between neighboring targets, of the order of 10 pc  in the Sun vicinity.}
              \label{FigGPstars}%
    \end{figure*}

\begin{figure*}[h!]
 \centering
 \includegraphics[width=12cm]{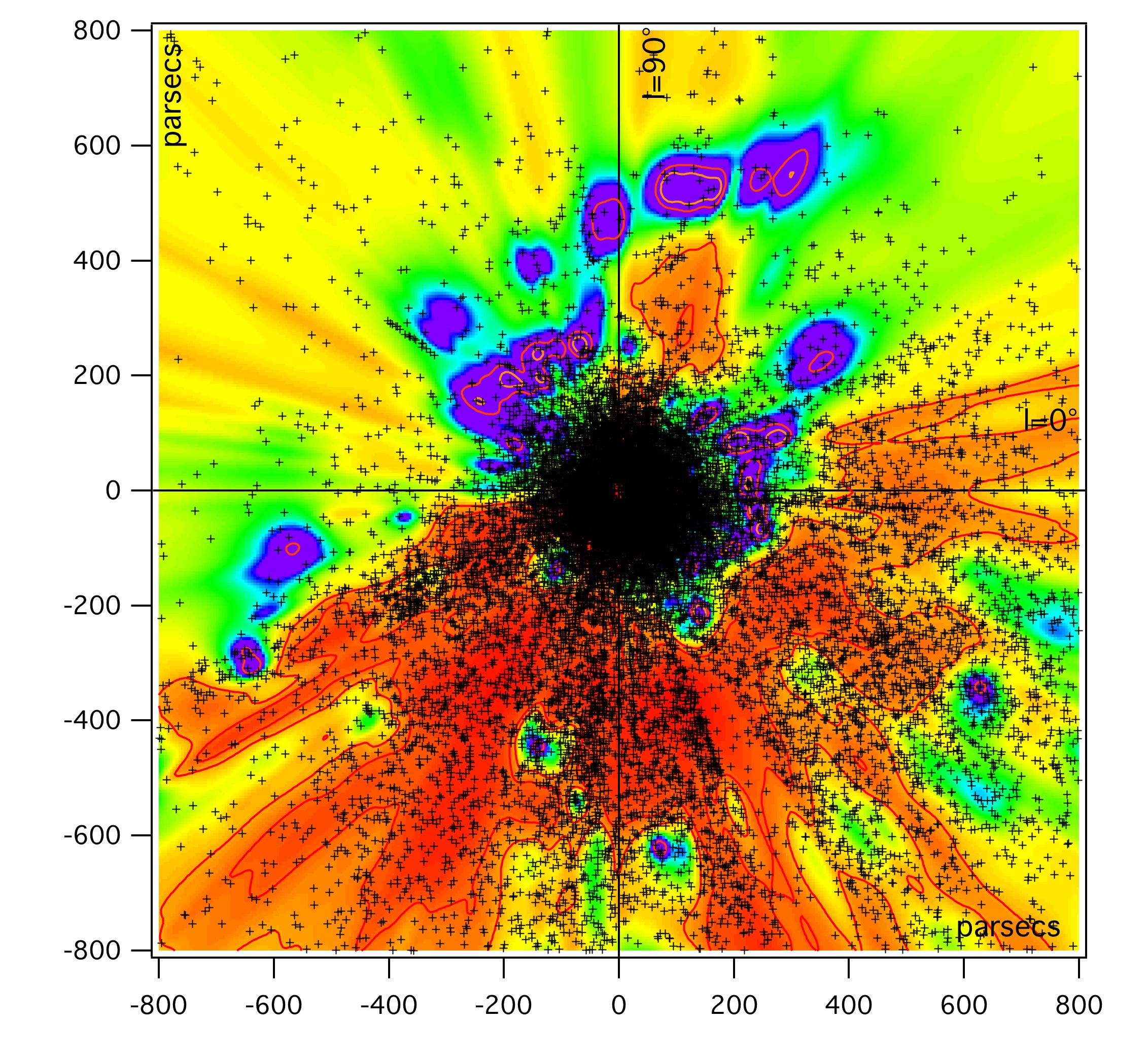}
 \caption{Same as Fig  \ref{FigGPstars}, but extending up to 800 pc and displaying stars within 150 pc from the Plane. The limiting size for the inverted structures varies from 40 to 200 pc at 800 pc depending on the regions.}
              \label{FigGPstars_extended}%
  \end{figure*}

The distribution of the target stars gives the constraints for the inversion. We show in Fig \ref{FigGPstars} those target stars that are located within 10 pc from the Plane and have mainly contributed to the computed distribution in  this plane, superimposed on the inner part of the computed map. It allows to infer the spatial resolution that can be achieved in a given location.
Fig \ref{FigGPstars_extended} shows those target stars that are located within 150 pc from the Plane and allows to figure out the achievable spatial resolution at larger distance.

\end{appendix}

\end{document}